\documentclass[journal]{IEEEtran}

\usepackage[table]{xcolor}
\usepackage{adjustbox}
\usepackage{algorithm}
\usepackage{algpseudocode}
\usepackage{amsfonts}
\usepackage{amsmath}
\usepackage{amssymb}
\usepackage{amsthm}
\usepackage{bookmark}
\usepackage{booktabs}
\usepackage[makeroom]{cancel}
\usepackage[american]{circuitikz}
\usepackage{cite}
\usepackage{fixmath}
\usepackage[acronym]{glossaries-extra}
\usepackage{hyperref}
\usepackage{import}
\usepackage{mathtools}
\usepackage{microtype}
\usepackage[short]{optidef}
\usepackage{pgfplots}
\usepackage{ragged2e}
\usepackage[subtle]{savetrees}
\usepackage{siunitx}
\usepackage{stfloats}
\usepackage[caption=false,font=footnotesize,subrefformat=parens,labelformat=parens]{subfig}
\usepackage{tabularx}
\usepackage{tikz}

\newtheorem{proposition}{Proposition}
\newtheorem{remark}{Remark}

\usetikzlibrary{arrows,calc,matrix,patterns,plotmarks,positioning,shapes}
\usetikzlibrary{decorations.pathmorphing,decorations.pathreplacing,decorations.shapes,shapes.geometric}
\usepgfplotslibrary{groupplots,patchplots}
\pgfplotsset{compat=newest}

\newcolumntype{L}{>{\RaggedRight}X}
\newcolumntype{C}{>{\centering\arraybackslash}X}

\makeatletter
\renewcommand{\fnum@algorithm}{\fname@algorithm{} \thealgorithm:}
\newcommand\setalgorithmcaptionfont[1]{%
	\let\my@floatc@ruled\floatc@ruled          
	\def\floatc@ruled{%
		\global\let\floatc@ruled\my@floatc@ruled 
		#1\floatc@ruled}}
\makeatother

\algrenewcommand{\algorithmicrequire}{\textbf{Input:}}
\algrenewcommand{\algorithmicensure}{\textbf{Output:}}
\algrenewcommand{\algorithmicwhile}{\textbf{While}}
\algrenewcommand{\algorithmicend}{\textbf{End}}
\algrenewcommand{\algorithmicrepeat}{\textbf{Repeat}}
\algrenewcommand{\algorithmicuntil}{\textbf{Until}}
\algrenewcommand{\algorithmicdo}{}

\glsdisablehyper
\setabbreviationstyle[acronym]{long-short}
\newacronym{af}{AF}{Amplify-and-Forward}
\newacronym{ambc}{AmBC}{Ambient Backscatter Communication}
\newacronym{ap}{AP}{Access Point}
\newacronym{awgn}{AWGN}{Additive White Gaussian Noise}
\newacronym{bcd}{BCD}{Block Coordinate Descent}
\newacronym{bc}{BackCom}{Backscatter Communication}
\newacronym{bibo}{BIBO}{Binary-Input Binary-Output}
\newacronym{bpcu}{\si{bpcu}}{bits per channel use}
\newacronym{bpsphz}{\si{bps/Hz}}{bits per second per Hertz}
\newacronym{clt}{CLT}{Central Limit Theorem}
\newacronym{cw}{CW}{Continuous Waveform}
\newacronym{cp}{CP}{Canonical Polyadic}
\newacronym{cr}{CR}{Cognitive Radio}
\newacronym{cscg}{CSCG}{Circularly Symmetric Complex Gaussian}
\newacronym{csi}{CSI}{Channel State Information}
\newacronym{dc}{DC}{Direct Current}
\newacronym{df}{DF}{Decode-and-Forward}
\newacronym{dmc}{DMC}{Discrete Memoryless Channel}
\newacronym{dmtc}{DMTC}{Discrete Memoryless Thresholding Channel}
\newacronym{dmmac}{DMMAC}{Discrete Memoryless Multiple Access Channel}
\newacronym{dcmc}{DCMC}{Discrete-input Continuous-output Memoryless Channel}
\newacronym{dp}{DP}{Dynamic Programming}
\newacronym{fdma}{FDMA}{Frequency-Division Multiple Access}
\newacronym{iid}{i.i.d.}{independent and identically distributed}
\newacronym{ioe}{IoE}{Internet of Everything}
\newacronym{iot}{IoT}{Internet of Things}
\newacronym{kkt}{KKT}{Karush-Kuhn-Tucker}
\newacronym{m2m}{M2M}{Machine to Machine}
\newacronym{mac}{MAC}{Multiple Access Channel}
\newacronym{mc}{MC}{Multiplication Coding}
\newacronym{miso}{MISO}{Multiple-Input Single-Output}
\newacronym{mimo}{MIMO}{Multiple-Input Multiple-Output}
\newacronym{ml}{ML}{Maximum-Likelihood}
\newacronym{mrt}{MRT}{Maximum Ratio Transmission}
\newacronym{noma}{NOMA}{Non-Orthogonal Multiple Access}
\newacronym{ofdm}{OFDM}{Orthogonal Frequency-Division Multiplexing}
\newacronym{pdf}{PDF}{Probability Density Function}
\newacronym{pga}{PGA}{Projected Gradient Ascent}
\newacronym{psk}{PSK}{Phase Shift Keying}
\newacronym{qam}{QAM}{Quadrature Amplitude Modulation}
\newacronym{qos}{QoS}{Quality of Service}
\newacronym{rf}{RF}{Radio-Frequency}
\newacronym{rfid}{RFID}{Radio-Frequency Identification}
\newacronym{ris}{RIS}{Reconfigurable Intelligent Surface}
\newacronym{sc}{SC}{Superposition Coding}
\newacronym{sic}{SIC}{Successive Interference Cancellation}
\newacronym{simo}{SIMO}{Single-Input Multiple-Output}
\newacronym{sinr}{SINR}{Signal-to-Interference-plus-Noise Ratio}
\newacronym{smawk}{SMAWK}{Shor-Moran-Aggarwal-Wilber-Klawe}
\newacronym{snr}{SNR}{Signal-to-Noise Ratio}
\newacronym{sr}{SR}{Symbiotic Radio}
\newacronym{swipt}{SWIPT}{Simultaneous Wireless Information and Power Transfer}
\newacronym{tdma}{TDMA}{Time-Division Multiple Access}
\newacronym{ue}{UE}{user}
\newacronym{wit}{WIT}{Wireless Information Transfer}
\newacronym{wpcn}{WPCN}{Wireless Powered Communication Network}
\newacronym{wpt}{WPT}{Wireless Power Transfer}
\newacronym{mbc}{MBC}{Monostatic \glsentryshort{bc}}
\newacronym{bbc}{BBC}{Bistatic \glsentryshort{bc}}
\newacronym{bls}{BLS}{Backtracking Line Search}
\newacronym{mrc}{MRC}{Maximal Ratio Combining}
\newacronym{sdma}{SDMA}{Space-Division Multiple Access}
\newacronym{nlos}{NLoS}{Non-Line-of-Sight}
\newacronym{zf}{ZF}{Zero-Forcing}
\newacronym{mmse}{MMSE}{Minimum Mean-Square-Error}
\newacronym{fpga}{FPGA}{Field-Programmable Gate Array}
\newacronym{ber}{BER}{Bit Error Rate}

\begin{document}
\title{RIScatter: Unifying Backscatter Communication and Reconfigurable Intelligent Surface}
\author{
	\IEEEauthorblockN{
		Yang~Zhao,~\IEEEmembership{Member,~IEEE,}
		and~Bruno~Clerckx,~\IEEEmembership{Fellow,~IEEE}
	}
	\thanks{
		The authors are with the Department of Electrical and Electronic Engineering, Imperial College London, London SW7 2AZ, U.K. (e-mail: \{yang.zhao18, b.clerckx\}@imperial.ac.uk).
		B. Clerckx is also with Silicon Austria Labs (SAL), Graz A-8010, Austria.
	}
}
\maketitle

\begin{abstract}
	\gls{bc} nodes harvest energy from and modulate information over external carriers.
	\gls{ris} adapts phase shift response to alter channel strength in specific directions.
	In this paper, we unify those two seemingly different technologies (and their derivatives) into one architecture called RIScatter.
	RIScatter is a batteryless cognitive radio that recycles ambient signal in an adaptive and customizable manner, where dispersed or co-located scatter nodes partially modulate their information and partially engineer the wireless channel.
	The key is to render the probability distribution of reflection states as a joint function of the information source, \gls{csi}, and relative priority of coexisting links.
	This enables RIScatter to softly bridge \gls{bc} and \gls{ris}; reduce to either in special cases; or evolve in a mixed form for heterogeneous traffic control and universal hardware design.
	We also propose a low-complexity \gls{sic}-free receiver that exploits the properties of RIScatter.
	For a single-user multi-node network, we characterize the achievable primary-(total-)backscatter rate region by optimizing the input distribution at scatter nodes, the active beamforming at the \gls{ap}, and the energy decision regions at the user.
	Simulations demonstrate RIScatter nodes can shift between backscatter modulation and passive beamforming.
\end{abstract}


\glsresetall

\begin{section}{Introduction}
	\label{sc:introduction}
	\IEEEPARstart{F}{uture} wireless network is envisioned to provide high throughput, uniform coverage, pervasive connectivity, heterogeneous control, and cognitive intelligence for trillions of low-power devices.
	\gls{bc} separates a transmitter into a \gls{rf} carrier emitter with power-hungry elements (e.g., synthesizer and amplifier) and an information-bearing node with power-efficient components (e.g., harvester and modulator) \cite{Boyer2014}.
	The receiver (reader) can be either co-located or separated with the carrier emitter, known as \gls{mbc} and \gls{bbc} in Fig.~\subref*{fg:mbc} and \subref*{fg:bbc}, respectively.
	Relevant applications such as \gls{rfid} \cite{Dobkin2012,Landt2005} and passive sensor network \cite{Vannucci2008,Assimonis2016} have been extensively researched, standardized, and commercialized to embrace the \gls{ioe}.
	However, conventional backscatter nodes only respond when externally inquired by a nearby reader.
	\gls{ambc} in Fig.~\subref*{fg:ambc} was proposed a decade ago where battery-free nodes recycle ambient signals (e.g., radio, television and Wi-Fi) to harvest energy and establish connections \cite{Liu2013b}.
	It does not require dedicated power source, carrier emitter, or frequency spectrum, but the backscatter decoding is subject to the strong interference from the primary (legacy) link.
	To tackle this, cooperative \gls{ambc} \cite{Yang2018} employs a co-located receiver to decode both coexisting links, and the concept was further refined as \gls{sr} in Fig.~\subref*{fg:sr} \cite{Liang2020}.
	Specifically, the active transmitter generates \gls{rf} wave carrying primary information, the passive node creates a rich-scattering environment and superimposes its own information, and the co-located receiver cooperatively decodes both links.
	In those \gls{bc} applications, the scatter node is considered as an \emph{information source} and the reflection pattern depends exclusively on the information symbol.
	On the other hand, \gls{ris} in Fig.~\subref*{fg:ris} is a smart signal reflector with numerous passive elements of adjustable phase shifts.
	It customizes the wireless environment for signal enhancement, interference suppression, scattering enrichment, and/or non-line-of-sight bypassing \cite{Wu2021b}.
	Each \gls{ris} element is considered as a \emph{channel shaper} and the reflection pattern depends exclusively on the \gls{csi}.

	As a special case of \gls{cr}, active and passive transmissions coexist and interplay in \gls{ambc} and \gls{sr}.
	Such a coexistence is classified into commensal (overlay), parasitic (underlay), and competitive (interfering) paradigms, and their achievable rate and outage performance were investigated in \cite{Guo2019b,Ding2020}.
	The achievable rate and optimal input distribution for binary-input \gls{ambc} were investigated in \cite{Qian2019b}, but its impact on the primary link was omitted.
	In \cite{Hassani2023}, the authors analyzed the energy efficiency and achievable rate region for an \gls{ambc}-aided multi-user downlink \gls{noma} system.
	However, they assumed equal symbol duration and perfect synchronization for the coexisting links.
	Importantly, active-passive coexisting networks have three special and important properties:
	\begin{enumerate}
		\item Primary and backscatter symbols are superimposed by \emph{double modulation} (i.e., multiplication coding);
		\item Backscatter signal strength is much weaker than primary due to \emph{double fading};
		\item The spreading factor (i.e., backscatter symbol duration over primary) is usually large\footnote{The load-switching interval of low-power backscatter modulators is usually \num{0.1} to \qty{10}{\us} \cite{Torres2021}, accounting for a typical spreading factor between \num{10} and \num{e3}.}.
	\end{enumerate}
	The second property motivated \cite{Long2020a,Liang2020,Guo2019b,Ding2020,Hassani2023,Zhou2019a,Wu2021a,Xu2021a,Yang2021a,Yang2018,Han2021,Zhang2022,Dai2023} to view \gls{sr} as a multiplicative \gls{noma} and perform \gls{sic} from primary to backscatter link.
	During primary decoding, the backscatter signal can be modelled as channel uncertainty or multiplicative interference, when the spreading factor is large or small.
	Decoding each backscatter symbol also requires multiple \gls{sic} followed by a \gls{mrc} over primary blocks, which is operation-intensive and \gls{csi}-sensitive.
	Under those assumptions, the achievable rate region of cell-free \gls{sr} was characterized in \cite{Dai2023}.
	When the spreading factor is sufficiently large, the primary achievable rate under semi-coherent detection\footnote{In this paper, semi-coherent detection refers to the primary/backscatter decoding with known \gls{csi} and unknown backscatter/primary symbols.} asymptotically approaches its coherent counterpart such that both links are approximately interference-free \cite{Long2020a}.
	However, this assumption severely limits the backscatter throughput.

	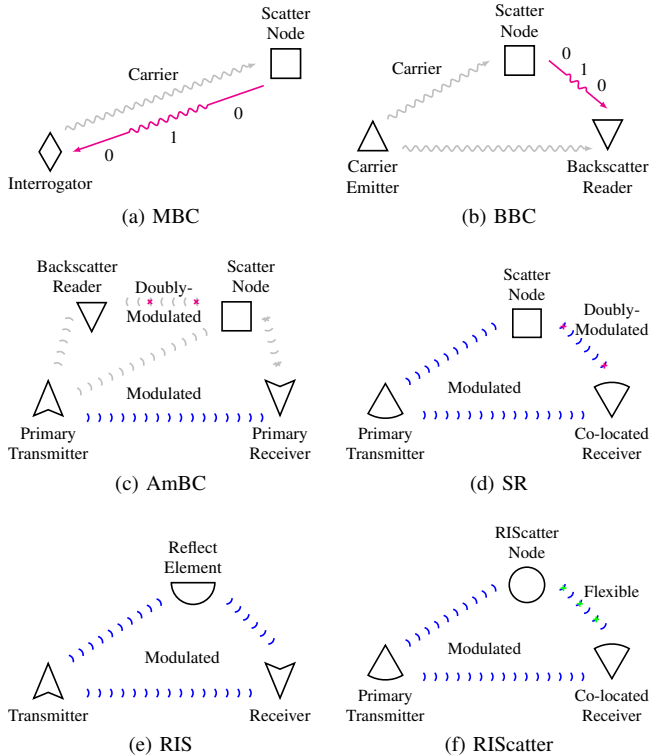
\begin{figure}[!t]
		\centering
		\subfloat[\gls{mbc}]{
			\resizebox{0.48\linewidth}{!}{
				\begin{tikzpicture}[font=\LARGE,every node/.style={draw,ultra thick},every path/.style={ultra thick},every text node part/.style={align=center}]
	\node at (0,0) [kite,minimum size=0.75cm,kite vertex angles=60] {};
	\node at (8,3) [rectangle,minimum size=1cm] {};

	\node[draw=none] at (0,-1.125) {Interrogator};
	\node[draw=none] at (8,4.375) {Scatter\\Node};

	\draw[lightgray,decorate,decoration={snake,post length=2.5mm},-latex] (0.5,0.75) -- (7,3);
	\draw[magenta,decorate,decoration={snake,pre length=2cm,post length=2cm},-latex] (7.25,2.25) -- (0.75,0);

	\node[draw=none] at (3.5,2.625) {Carrier};
	\node[draw=none] at (4.2,0.5) {1};
	\node[draw=none] at (2,-0.125) {0};
	\node[draw=none] at (6.4,1.3) {0};
\end{tikzpicture}
			}
			\label{fg:mbc}
		}
		\subfloat[\gls{bbc}]{
			\resizebox{0.48\linewidth}{!}{
				\begin{tikzpicture}[font=\LARGE,every node/.style={draw,ultra thick},every path/.style={ultra thick},every text node part/.style={align=center}]
	\node at (0,0) [isosceles triangle,isosceles triangle stretches,shape border rotate=+90,minimum size=1cm,minimum height=1cm,anchor=north] {};
	\node at (5,2) [rectangle,minimum size=1cm] {};
	\node at (8,0) [isosceles triangle,isosceles triangle stretches,shape border rotate=+270,minimum size=1cm,minimum height=1cm,anchor=north] {};

	\node[draw=none] at (0,-1.95) {Carrier\\Emitter};
	\node[draw=none] at (5,3.35) {Scatter\\Node};
	\node[draw=none] at (8,-1.95) {Backscatter\\Reader};

	\draw[lightgray,decorate,decoration={snake,post length=2.5mm},-latex] (1,-1) -- (7.5,-1);
	\draw[lightgray,decorate,decoration={snake,post length=2.5mm},-latex] (0.5,0) -- (4,2);
	\draw[magenta,decorate,decoration={snake,pre length=7.5mm,post length=7.5mm},-latex] (6,2) -- (8,0.35);

	\node[draw=none] at (1.5,1.75) {Carrier};
	\node[draw=none] at (7.2,1.7) {1};
	\node[draw=none] at (6.6,2.25) {0};
	\node[draw=none] at (7.8,1.15) {0};
\end{tikzpicture}
			}
			\label{fg:bbc}
		}
		\\
		\subfloat[\gls{ambc}]{
			\resizebox{0.48\linewidth}{!}{
				\begin{tikzpicture}[font=\LARGE,every node/.style={draw,ultra thick},every path/.style={ultra thick},every text node part/.style={align=center}]
	\node at (0,0) [dart,shape border rotate=90,minimum width=1cm] {};
	\node at (1.5,3.5) [isosceles triangle,isosceles triangle stretches,shape border rotate=+270,minimum size=1cm,minimum height=1cm,anchor=north] {};
	\node at (6.5,3) [rectangle,minimum size=1cm] {};
	\node at (8,0.35) [dart,shape border rotate=270,minimum width=1cm] {};

	\node[draw=none] at (0,-1.375) {Primary\\Transmitter};
	\node[draw=none] at (1,4.375) {Backscatter\\Reader};
	\node[draw=none] at (7,4.375) {Scatter\\Node};
	\node[draw=none] at (8,-1.375) {Primary\\Receiver};

	\draw[blue,decorate,decoration={waves,segment length=4mm,radius=2mm}] (1,-0.5) -- (7.5,-0.5);
	\draw[lightgray,decorate,decoration={waves,segment length=4mm,radius=2mm}] (0.25,1) -- (0.75,3);
	\draw[lightgray,decorate,decoration={waves,segment length=4mm,radius=2mm}] (0.75,0.25) -- (5.5,2.5);
	\draw[lightgray,decorate,decoration={waves,segment length=4mm,radius=2mm}] (5.5,3.5) -- (2.5,3.5);
	\draw[magenta,decorate,decoration={crosses,segment length=16mm,shape size=1.5mm}] (5.11,3.5) -- (2.5,3.5);
	\draw[lightgray,decorate,decoration={waves,segment length=4mm,radius=2mm}] (7.43,3.28) -- (8,1);
	\draw[lightgray,decorate,decoration={crosses,segment length=16mm,shape size=1.5mm}] (7.525,2.9) -- (8,1);

	\node[draw=none] at (4,0.35) {Modulated};
	\node[draw=none] at (4,3.95) {Doubly-};
	\node[draw=none] at (4,3) {Modulated};
\end{tikzpicture}
			}
			\label{fg:ambc}
		}
		\subfloat[\gls{sr}]{
			\resizebox{0.48\linewidth}{!}{
				\begin{tikzpicture}[font=\LARGE,every node/.style={draw,ultra thick},every path/.style={ultra thick},every text node part/.style={align=center}]
	\node at (0,0) [circular sector,shape border rotate=90,minimum width=2cm] {};
	\node at (5,3) [rectangle,minimum size=1cm] {};
	\node at (8,0.55) [circular sector,shape border rotate=270,minimum width=2cm] {};

	\node[draw=none] at (0,-1.25) {Primary\\Transmitter};
	\node[draw=none] at (5,4.375) {Scatter\\Node};
	\node[draw=none] at (8,-1.25) {Co-located\\Receiver};

	\draw[blue,decorate,decoration={waves,segment length=4mm,radius=2mm}] (1,-0.25) -- (7.375,-0.25);
	\draw[blue,decorate,decoration={waves,segment length=4mm,radius=2mm}] (0.5,0.75) -- (4,3);
	\draw[blue,decorate,decoration={waves,segment length=4mm,radius=2mm}] (6,3.1625) -- (8,1.25);
	\draw[magenta,decorate,decoration={crosses,segment length=20mm,shape size=1.5mm}] (6.29,2.885) -- (8,1.25);

	\node[draw=none] at (3.5,0.75) {Modulated};
	\node[draw=none] at (8,3.15) {Doubly-\\Modulated};
\end{tikzpicture}
			}
			\label{fg:sr}
		}
		\\
		\subfloat[\gls{ris}]{
			\resizebox{0.48\linewidth}{!}{
				\begin{tikzpicture}[font=\LARGE,every node/.style={draw,ultra thick},every path/.style={ultra thick},every text node part/.style={align=center}]
	\node at (0,0) [dart,shape border rotate=90,minimum width=1cm] {};
	\node at (5,3.25) [semicircle,shape border rotate=180,minimum width=1.5cm] {};
	\node at (8,0.35) [dart,shape border rotate=270,minimum width=1cm] {};

	\node[draw=none] at (0,-1) {Transmitter};
	\node[draw=none] at (5,4.4375) {Reflect\\Element};
	\node[draw=none] at (8,-1) {Receiver};

	\draw[blue,decorate,decoration={waves,segment length=4mm,radius=2mm}] (1,-0.25) -- (7.375,-0.25);
	\draw[blue,decorate,decoration={waves,segment length=4mm,radius=2mm}] (0.5,0.75) -- (4,3);
	\draw[blue,decorate,decoration={waves,segment length=4mm,radius=2mm}] (6,3.1625) -- (8,1.25);

	\node[draw=none] at (4.625,1) {Modulated};
\end{tikzpicture}
			}
			\label{fg:ris}
		}
		\subfloat[RIScatter]{
			\resizebox{0.48\linewidth}{!}{
				\begin{tikzpicture}[font=\LARGE,every node/.style={draw,ultra thick},every path/.style={ultra thick},every text node part/.style={align=center}]
	\node at (0,0) [circular sector,shape border rotate=90,minimum width=2cm] {};
	\node at (5,3) [circle,minimum size=1.25cm] {};
	\node at (8,0.55) [circular sector,shape border rotate=270,minimum width=2cm] {};

	\node[draw=none] at (0,-1.25) {Primary\\Transmitter};
	\node[draw=none] at (5,4.5) {RIScatter\\Node};
	\node[draw=none] at (8,-1.25) {Co-located\\Receiver};

	\draw[blue,decorate,decoration={waves,segment length=4mm,radius=2mm}] (1,-0.25) -- (7.375,-0.25);
	\draw[blue,decorate,decoration={waves,segment length=4mm,radius=2mm}] (0.5,0.75) -- (4,3);
	\draw[blue,decorate,decoration={waves,segment length=4mm,radius=2mm}] (6,3.1625) -- (8,1.25);
	\draw[green,decorate,decoration={crosses,segment length=8mm,shape size=1.5mm}] (6.29,2.885) -- (8,1.25);

	\node[draw=none] at (3.5,0.75) {Modulated};
	\node[draw=none] at (8,2.76) {Flexible};
\end{tikzpicture}
			}
			\label{fg:riscatter}
		}
		\caption{
			Illustration of scattering applications.
			The blue flow(s) constitutes the primary link while the magenta/green flow denotes the backscatter link.
		}
		\label{fg:scatter_illustration}
	\end{figure}

	\begin{table*}[!t]
		\caption{Comparison of Scattering Applications}
		\label{tb:scattering_applications}
		\rowcolors{2}{gray!25}{white}
		\renewcommand{\arraystretch}{1.4}
		\begin{tabularx}{\textwidth}{CCCCCC}
			\toprule
			\hiderowcolors
			                                       & \glsentryshort{mbc}/\glsentryshort{bbc} & \glsentryshort{ambc}        & \glsentryshort{sr} (large spreading factor)                         & \glsentryshort{ris} & RIScatter                                                          \\ \midrule
			\showrowcolors
			Information link(s)                    & Backscatter                             & Coexisting                  & Coexisting                                                          & Primary             & Coexisting                                                         \\
			Primary signal on backscatter decoding & Carrier                                 & Multiplicative interference & Spreading code                                                      & ---                 & Energy uncertainty                                                 \\
			Backscatter signal on primary decoding & ---                                     & Multiplicative interference & \glsentryshort{csi} uncertainty                                     & Passive beamforming & Dynamic passive beamforming                                        \\
			Cooperative devices                    & ---                                     & No                          & Primary transmitter and co-located receiver                         & ---                 & Primary transmitter, scatter nodes, and co-located receiver        \\
			Sequential decoding                    & ---                                     & No                          & Primary-to-backscatter, \glsentryshort{sic} and \glsentryshort{mrc} & ---                 & Backscatter-to-primary, no \glsentryshort{sic}/\glsentryshort{mrc} \\
			Reflection pattern depends on          & Information source                      & Information source          & Information source                                                  & \glsentryshort{csi} & Information source, \glsentryshort{csi}, and \glsentryshort{qos}   \\
			Reflection state distribution          & Equiprobable                            & Equiprobable                & Equiprobable or Gaussian                                            & Degenerate          & Flexible                                                           \\
			Load-switching speed                   & Fast                                    & Slow                        & Slow                                                                & Quasi-static        & Arbitrary                                                          \\ \bottomrule
		\end{tabularx}
	\end{table*}

	On the other hand, static \gls{ris} that employs fixed reflection pattern per channel block has been extensively studied in wireless communication, sensing, and power literature \cite{Wu2018,Zhang2019a,Lin2022,Liu2022,Feng2022,Zhao2022}.
	Dynamic \gls{ris} performs time sharing between different phase shifts and introduces artificial channel diversity within each channel block.
	The idea was first proposed to fine-tune the \gls{ofdm} resource blocks \cite{Yang2020}, then extended to the downlink power and uplink information phases of \gls{wpcn} \cite{Wu2021,Wu2021d,Hua2022a}.
	However, dynamic \gls{ris} carries no information because the reflection state at a specific time is known to the receiver.
	\gls{ris} can also be used as an information source, and prototypes have been developed for \gls{psk} \cite{Tang2019a} and \gls{qam} \cite{Dai2020a}.
	From an information-theoretic perspective, the authors of \cite{Karasik2020} reported that joint transmitter-\gls{ris} encoding achieves the capacity of \gls{ris}-aided finite-input channel, and using \gls{ris} as a naive passive beamformer to maximize the receive \gls{snr} is generally rate-suboptimal.
	This inspired \cite{Liu2019d,Bereyhi2020,Xu2020b,Zhang2021d,Hu2021b,Hua2022,Basar2020,Ma2020a,Yuan2021,Hu2021a} to combine passive beamforming and backscatter modulation in the overall \gls{ris} design.
	In particular, \emph{symbol level precoding} maps the information symbols to the optimized \gls{ris} coefficient sets \cite{Liu2019d,Bereyhi2020}, \emph{overlay modulation} superposes the information symbols over a common auxiliary matrix \cite{Xu2020b,Zhang2021d,Hu2021b,Hua2022}, \emph{spatial modulation} switches between the reflection coefficient sets that maximize \gls{snr} at different receive antennas \cite{Basar2020,Ma2020a,Yuan2021}, and \emph{index modulation} employs dedicated reflection elements (resp. information elements) for passive beamforming (resp. backscatter modulation) \cite{Hu2021a}.
	Those \gls{ris}-based backscatter modulation schemes incur advanced hardware architecture and high optimization complexity.
	In contrast, \cite{Vardakis2023} exploited commodity \gls{rfid} tags, powered and controlled by a software-defined radio reader at a different frequency, to perform passive beamforming (but no backscatter modulation) towards a legacy user. \label{cm:2.5}
	Most relevant literature considered either Gaussian codebook \cite{Guo2019b,Ding2020,Long2020a,Zhou2019a,Wu2021a,Xu2021a,Yang2021a,Hu2021b} that is impractical for low-power nodes, or finite equiprobable inputs \cite{Yang2018,Liang2020,Han2021,Zhang2022,Liu2019d,Bereyhi2020,Xu2020b,Zhang2021d,Hua2022,Basar2020,Ma2020a,Yuan2021,Hu2021a} that does not fully exploit the \gls{csi} and properties of active-passive coexisting networks.
	Those problems are addressed in this paper and the contributions are summarized below.

	\emph{First,} we propose RIScatter as a novel protocol that unifies \gls{bc} and \gls{ris} by adaptive reflection state (backscatter input) distribution design.
	The concept is shown in Fig.~\subref*{fg:riscatter}, where one or more RIScatter nodes ride over an active transmission to simultaneously modulate their information and engineer the wireless channel.
	A co-located receiver cooperatively decodes both coexisting links.
	Each reflection state is simultaneously a passive beamforming codeword and part of information codeword.
	The reflection pattern over time is semi-random and guided by the input probability assigned to each state.
	This probability distribution is carefully designed to incorporate the backscatter information, \gls{csi}, and \gls{qos}\footnote{\gls{qos} refers to the relative priority of the primary link.\label{fn:qos}}.
	Such an adaptive channel coding boils down to the degenerate distribution of \gls{ris} when the primary link is prioritized, and outperforms the uniform distribution of \gls{bc} (by accounting the \gls{csi}) when the backscatter link is prioritized.
	Table~\ref{tb:scattering_applications} compares RIScatter to \gls{bc} and \gls{ris}.
	However, two major challenges for RIScatter are the receiver design and input distribution design.
	This is the first paper to unify \gls{bc} and \gls{ris} from the perspective of input distribution.

	\emph{Second,} we address the first challenge and propose a low-complexity \gls{sic}-free receiver.
	It semi-coherently decodes the weak backscatter signal using an energy detector, re-encodes for the exact reflection pattern, then coherently decodes the primary link.
	Thanks to double modulation, backscatter detection can be viewed as part of channel training, and the impact of backscatter modulation can be modelled as dynamic passive beamforming afterwards.
	The proposed receiver may be built over legacy receivers with minor hardware upgrade, as it only requires one additional energy comparison and re-encoding per backscatter symbol (instead of primary symbol).
	The energy detector can also be tailored for arbitrary input distribution and spreading factor to increase backscatter throughput.
	This is the first paper to propose a \gls{sic}-free cooperative receiver for active-passive coexisting networks.

	\emph{Third,} we address the second challenge in a single-user multi-node \gls{miso} scenario.
	We characterize the achievable primary-(total-)backscatter rate region by optimizing the input distribution at RIScatter nodes, the active beamforming at the \gls{ap}, and the energy decision regions at the user under different \gls{qos}.
	A \gls{bcd} algorithm is proposed where the \gls{kkt} input distribution is numerically evaluated by the converging point of a sequence, the active beamforming is optimized by \gls{pga}, and the decision regions are refined by state-of-the-art sequential quantizer designs for \gls{dmtc}.
	Uniquely, our optimization problem takes into account the \gls{csi}, \gls{qos}, and backscatter constellation, and the resulting input distribution is applicable to other detection schemes.
	This is also the first paper to reveal the importance of backscatter input distribution and decision region designs in active-passive coexisting networks.


	\emph{Notations:}
	Italic, bold lower-case, and bold upper-case letters denote scalars, vectors and matrices, respectively.
	$\boldsymbol{0}$ and $\boldsymbol{1}$ denote zero and one array of appropriate size, respectively. $\mathbb{I}^{x \times y}$, $\mathbb{R}_+^{x \times y}$, and $\mathbb{C}^{x \times y}$ denote the unit, real nonnegative, and complex spaces of dimension $x \times y$, respectively.
	$j$ denotes the imaginary unit.
	$\mathrm{diag}(\cdot)$ returns a square matrix with the input vector on its main diagonal and zeros elsewhere.
	$\mathrm{card}(\cdot)$ returns the cardinality of a set.
	$\log(\cdot)$ denotes logarithm of base $e$.
	$(\cdot)^*$, $(\cdot)^\mathsf{T}$, $(\cdot)^\mathsf{H}$, $\lvert{\cdot}\rvert$, and $\lVert{\cdot}\rVert$ denote the conjugate, transpose, conjugate transpose (Hermitian), absolute value, and Euclidean norm operators, respectively.
	$(\cdot)^{(r)}$ and $(\cdot)^{\star}$ denote the $r$-th iterated and optimal results, respectively.
	The distribution of a \gls{cscg} random variable with zero mean and variance $\sigma^2$ is denoted by $\mathcal{CN}(0,\sigma^2)$, and $\sim$ means ``distributed as''.
\end{section}

\begin{section}{RIScatter}
	\label{sc:riscatter}
	\begin{subsection}{Principles}
		\label{sc:principles}
		\gls{rf} wave scattering or reflecting are often manipulated by passive antennas or programmable metamaterial \cite{Liang2022}.
		The former receives the impinging signals and reradiates some back to the space, while the latter reflects at the space-cell boundary and mainly applies a phase shift.
		In the scattered signal, the structural mode component depends on the scatterer geometry and material.
		Its impact is usually modelled as part of environment multipath \cite{Thomas2012,Liang2020}, or simply a baseband \gls{dc} offset when the impinging signal is a \gls{cw} \cite{Boyer2014}.
		On the other hand, the antenna mode component depends on the impedance mismatch and is widely exploited in scattering applications.
		For an antenna (resp. metamaterial) scatterer with $M$ reflection states, the reflection coefficient at state $m \in \mathcal{M} \triangleq \{1,\ldots,M\}$ is \cite{VanHuynh2017,Liang2022}
		\begin{equation}
			\Gamma_m = \frac{Z_m - Z^*}{Z_m + Z},
			\label{eq:reflection_coefficient}
		\end{equation}
		where $Z_m$ is the antenna load (resp. metamaterial cell) impedance at state $m$, and $Z$ is the antenna input (resp. medium characteristic) impedance.
		Specifically,
		\begin{itemize}
			\item \gls{bc}: The scatterer is an information source with random reflection pattern over time.
			The reflection coefficient is used merely as part of information codeword \cite{Thomas2012a}
			\begin{equation}
				\Gamma_m = \alpha_m \frac{c_m}{\max_{m'} \lvert c_{m'} \rvert},
				\label{eq:backscatter_modulation}
			\end{equation}
			where $\alpha_m \in \mathbb{I}$ is the amplitude scattering ratio at state $m$, and $c_m$ is the corresponding constellation point.
			\item \gls{ris}: The scatterer is a channel shaper with deterministic reflection pattern over time.
			The reflection coefficient is used merely as a passive beamforming codeword \cite{Wu2018}
			\begin{equation}
				\Gamma_m = \alpha_m \exp(j \theta_m),
				\label{eq:passive_beamforming}
			\end{equation}
			where $\theta_m$ is the phase shift at state $m$.\footnote{Most papers assume $\alpha_m = \alpha$, with $\alpha \ll 1$ for \gls{bc} and $\alpha=1$ for \gls{ris}.\label{fn:scattering_ratio}}
		\end{itemize}

		RIScatter generalizes \gls{bc} and \gls{ris} from a probabilistic perspective.
		Each reflection coefficient simultaneously acts as a passive beamforming codeword and part of information codeword.
		As shown in Fig.~\ref{fg:scatter_comparison}, the reflection pattern of each RIScatter node over time is semi-random and guided by the input probability assigned to each state.
		This probability distribution is carefully designed to incorporate the backscatter information, \gls{csi}, and \gls{qos}, in order to strike a balance between backscatter modulation and passive beamforming.
		\begin{figure}[!t]
			\centering
			\subfloat[Input Distribution]{
				\resizebox{0.8\columnwidth}{!}{
					\begin{tikzpicture}
	\begin{groupplot}
		[group style={
					rows=3,
					group name=plots,
					x descriptions at=edge bottom,
					y descriptions at=edge left,
					vertical sep=10
				},
			ybar,
			xmin=0,
			xmax=4,
			xtick={1,2,3,4},
			ymin=0,
			ymax=1,
			xlabel={Reflection State},
			enlarge x limits={value=0.25,upper},
			enlarge y limits={value=0.1,upper},
			width=10cm,
			height=2.6cm,
			every axis plot/.append style={bar shift=0}
		]
		\nextgroupplot
		\addplot[blue,fill=blue!30!white] coordinates {(-1,-1)};
		\addplot[blue,fill=blue!30!white,postaction={pattern=north west lines},pattern color=.] coordinates {(1,0.25)};
		\addplot[blue,fill=blue!30!white,postaction={pattern=north east lines},pattern color=.] coordinates {(2,0.25)};
		\addplot[blue,fill=blue!30!white,postaction={pattern=vertical lines},pattern color=.] coordinates {(3,0.25)};
		\addplot[blue,fill=blue!30!white,postaction={pattern=horizontal lines},pattern color=.] coordinates {(4,0.25)};
		\legend{BackCom}
		\nextgroupplot[ylabel={Probability Distribution}]
		\addplot[red,fill=red!30!white] coordinates {(-1,-1)};
		\addplot[red,fill=red!30!white,postaction={pattern=north west lines},pattern color=.] coordinates {(1,0)};
		\addplot[red,fill=red!30!white,postaction={pattern=north east lines},pattern color=.] coordinates {(2,1)};
		\addplot[red,fill=red!30!white,postaction={pattern=vertical lines},pattern color=.] coordinates {(3,0)};
		\addplot[red,fill=red!30!white,postaction={pattern=horizontal lines},pattern color=.] coordinates {(4,0)};
		\legend{RIS}
		\nextgroupplot
		\addplot[brown,fill=brown!30!white] coordinates {(-1,-1)};
		\addplot[brown,fill=brown!30!white,postaction={pattern=north west lines},pattern color=.] coordinates {(1,0.25)};
		\addplot[brown,fill=brown!30!white,postaction={pattern=north east lines},pattern color=.] coordinates {(2,0.75)};
		\addplot[brown,fill=brown!30!white,postaction={pattern=vertical lines},pattern color=.] coordinates {(3,0)};
		\addplot[brown,fill=brown!30!white,postaction={pattern=horizontal lines},pattern color=.] coordinates {(4,0)};
		\legend{RIScatter}
	\end{groupplot}
\end{tikzpicture}
				}
				\label{fg:input_distribution}
			}
			\\
			\subfloat[Reflection Pattern]{
				\resizebox{0.8\columnwidth}{!}{
					\begin{tikzpicture}[font=\small]
	\foreach \x in {0}{
			\draw[blue,ultra thick,postaction={pattern=north west lines},pattern color=.] (2*\x,4) rectangle ++(2,0.25);
		}
	\foreach \x in {1}{
			\draw[blue,ultra thick,postaction={pattern=north east lines},pattern color=.] (2*\x,4) rectangle ++(2,0.25);
		}
	\foreach \x in {2}{
			\draw[blue,ultra thick,postaction={pattern=vertical lines},pattern color=.] (2*\x,4) rectangle ++(2,0.25);
		}
	\foreach \x in {3}{
			\draw[blue,ultra thick,postaction={pattern=horizontal lines},pattern color=.] (2*\x,4) rectangle ++(2,0.25);
		}
	\draw[latex-latex] (0,4.35) -- (2,4.35) node[above,midway]{BB};
	\node at (-1.5,4.125) {MBC/BBC};

	\foreach \x in {0,...,7}{
			\draw[teal,fill=teal!35!white,postaction={pattern=north west lines},pattern color=.] (0.5*\x,3) rectangle ++(0.5,0.25);
		}
	\foreach \x in {8,...,15}{
			\draw[teal,fill=teal!35!white,postaction={pattern=north east lines},pattern color=.] (0.5*\x,3) rectangle ++(0.5,0.25);
		}
	\foreach \x in {0,...,1}{
			\draw[teal,ultra thick] (4*\x,3) rectangle ++(4,0.25);
		}
	\draw[latex-latex] (0,3.35) -- (4,3.35) node[above,midway]{BB};
	\draw[latex-latex] (0,2.9) -- (0.5,2.9) node[below,midway]{PB};
	\node at (-1.5,3.125) {AmBC/SR};

	\foreach \x in {0,...,15}{
			\draw[red,fill=red!35!white,postaction={pattern=north east lines},pattern color=.] (0.5*\x,2) rectangle ++(0.5,0.25);
		}
	\draw[red,ultra thick] (0,2) rectangle ++(8,0.25);
	\draw[latex-latex] (0,2.35) -- (8,2.35) node[above,midway]{CB};
	\draw[latex-latex] (0,1.9) -- (0.5,1.9) node[below,midway]{PB};
	\node at (-1.5,2.125) {RIS};

	\foreach \x in {0,...,3}{
			\draw[brown,fill=brown!35!white,postaction={pattern=north west lines},pattern color=.] (0.5*\x,1) rectangle ++(0.5,0.25);
		}
	\foreach \x in {4,...,15}{
			\draw[brown,fill=brown!35!white,postaction={pattern=north east lines},pattern color=.] (0.5*\x,1) rectangle ++(0.5,0.25);
		}
	\foreach \x in {0,...,3}{
			\draw[brown,ultra thick] (2*\x,1) rectangle ++(2,0.25);
		}
	\draw[latex-latex] (0,1.35) -- (2,1.35) node[above,midway]{BB};
	\draw[latex-latex] (0,0.9) -- (0.5,0.9) node[below,midway]{PB};

	\node at (-1.5,1.125) {RIScatter};

	\draw [pattern=north west lines,ultra thick](0.15,5) rectangle ++(0.5,0.25) node[xshift=20,yshift=-3.5] {State 1};
	\draw [pattern=north east lines,ultra thick](2.15,5) rectangle ++(0.5,0.25) node[xshift=20,yshift=-3.5] {State 2};
	\draw [pattern=vertical lines,ultra thick](4.15,5) rectangle ++(0.5,0.25) node[xshift=20,yshift=-3.5] {State 3};
	\draw [pattern=horizontal lines,ultra thick](6.15,5) rectangle ++(0.5,0.25) node[xshift=20,yshift=-3.5] {State 4};
\end{tikzpicture}
				}
				\label{fg:time_structure}
			}
			\caption{
				Input distribution and reflection pattern of scattering applications.
				``PB'', ``BB'', and ``CB'' refer to primary symbol block, backscatter symbol block, and channel block, respectively.
				Shadowing means presence of primary link.
				In this example, the optimal passive beamformer corresponds to state \num{2}.
				The spreading factor is \num{4} for RIScatter and \num{8} for \gls{ambc}/\gls{sr}.
				\gls{bc} and \gls{ris} can be viewed as extreme cases of RIScatter, where the input distribution boils down to uniform and degenerate, respectively.
			}
			\label{fg:scatter_comparison}
		\end{figure}
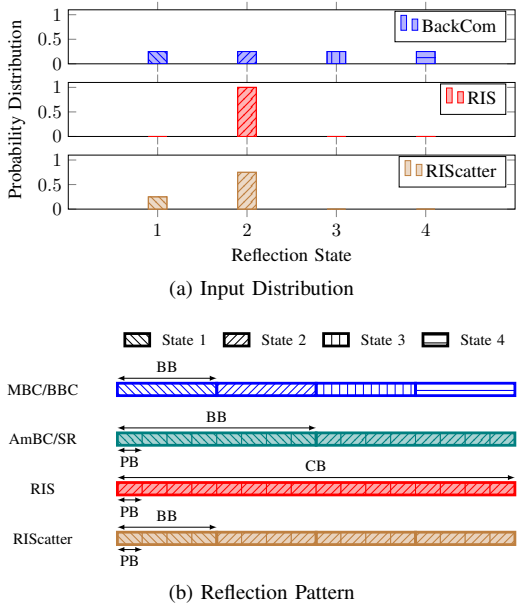
		\begin{remark}
			Unlike dynamic \gls{ris} that simply performs a time sharing between reflection states, RIScatter conveys additional information by randomizing the reflection pattern over time while still guaranteeing the probability of occurrence of each state.
			Upon successful backscatter detection, the impact of RIScatter nodes on the primary link can be modelled as dynamic passive beamforming.
			\label{re:dynamic_passive_beamforming}
		\end{remark}

		RIScatter nodes can be implemented, for example, by adding an integrated receiver\footnote{The aim is to coordinate the node with the active source and to acquire the optimized input distribution, instead of decoding the primary information dedicated for the user. The node receiver can be implemented using simple circuits or even integrated with the rectifier \cite{Kim2021a} to reduce cost and complexity.\label{fn:integrated_receiver}} \cite{Kim2021a} and adaptive encoder \cite{He2020e} to off-the-shelf passive \gls{rfid} tags.
		The block diagram, equivalent circuit, and scatter model are illustrated in Fig.~\ref{fg:riscatter_node}.
		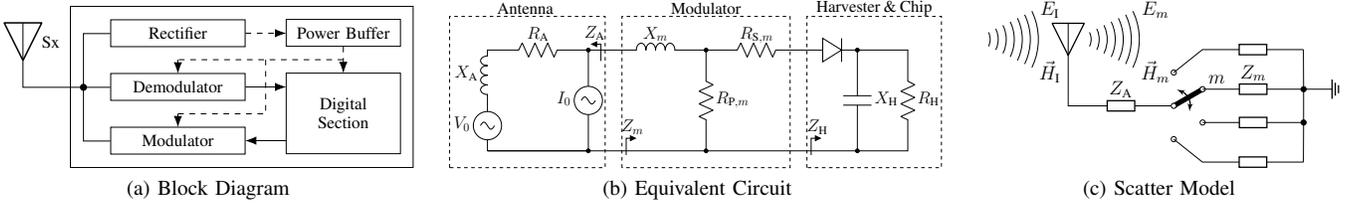
\begin{figure*}[!t]
			\centering
			\subfloat[Block Diagram]{
				\resizebox{0.32\linewidth}{!}{
					\makeatletter
\tikzset{
	block/.style={draw,rectangle,align=center},
	from/.style args={#1 to #2}{
			above right={0cm of #1},
			/utils/exec=\pgfpointdiff
			{\tikz@scan@one@point\pgfutil@firstofone(#1)\relax}
			{\tikz@scan@one@point\pgfutil@firstofone(#2)\relax},
			minimum width/.expanded=\the\pgf@x,
			minimum height/.expanded=\the\pgf@y
		}
}
\makeatother

\begin{circuitikz}[transform shape]
	\tikzstyle{every node}=[font=\small]
	\coordinate (O) at (0,0);
	\node[block,from={O to $(O) + (6.375,3)$}](T){};
	\draw (0,1.5)
	to[short] ++(-0.875,0)
	to[short] ++(0,0.5) node[bareantenna](A){Sx};
	\draw (0,1.5)
	to[short,-*] ++(0.25,0) coordinate(J);
	\draw (J)
	to[short] ++(0,1)
	to[short] ++(0.5,0) coordinate(J1);
	\node[block,from={$(J1) + (0,-0.25)$ to $(J1) + (2.5,0.25)$}](R){Rectifier};
	\draw (J)
	to[short] ++(0.5,0) coordinate(J2);
	\node[block,from={$(J2) + (0,-0.25)$ to $(J2) + (2.5,0.25)$}](D){Demodulator};
	\draw (J)
	to[short] ++(0,-1)
	to[short] ++(0.5,0) coordinate(J3);
	\node[block,from={$(J3) + (0,-0.25)$ to $(J3) + (2.5,0.25)$}](M){Modulator};
	\draw[dashed,-{Latex[length=2mm]}] (R.east) -- ++(0.75,0);
	\draw[-{Latex[length=2mm]}] (D.east) -- ++(0.75,0);
	\draw[{Latex[length=2mm]}-] (M.east) -- ++(0.75,0);
	\node[block,from={$(R.east) + (0.75,-0.25)$ to $(R.east) + (2.875,0.25)$}](P){Power Buffer};
	\node[block,from={$(M.east) + (0.75,-0.25)$ to $(D.east) + (2.875,0.25)$}](S){Digital\\Section};
	\draw[dashed,-{Latex[length=2mm]}] (P.south) to (S.north);
	\coordinate (F1) at ($(P.south)!0.5!(S.north)$);
	\coordinate (F2) at ($(D.east)!0.5!(S.west)$);
	\coordinate (F3) at ($(D.south)!0.5!(M.north)$);
	\draw[dashed] (F1) to (F1-|D.north);
	\draw[dashed,-{Latex[length=2mm]}] (F1-|D.north) to (D.north);
	\draw[dashed] (F1-|F2) to (F2|-F3) to (M|-F3);
	\draw[dashed,-{Latex[length=2mm]}] (M|-F3) to (M.north);
\end{circuitikz}
				}
				\label{fg:block_diagram}
			}
			\subfloat[Equivalent Circuit]{
				\resizebox{0.37\linewidth}{!}{
					\begin{circuitikz}[transform shape]
	\tikzstyle{every node}=[font=\Large]
	\draw (0,0) coordinate(O)
	to [sV,l=$V_0$] ++(0,1.5)
	to [L,l=$X_{\text{A}}$] ++(0,1.5)
	to [R=$R_{\text{A}}$,-*] ++(3,0) coordinate(AM)
	to [sI,l_=$I_0$,-*] ++(0,-3)
	to [short] (O);
	\draw (AM)
	to [short] ++(0.5,0)
	to [L,l=$X_m$,-*] ++(3,0) coordinate(M)
	to [R=$R_{\text{S},m}$] ++(3,0) coordinate(MH);
	\draw (M)
	to [R=$R_{\text{P},m}$,-*] ++(0,-3);
	\draw (MH)
	to [D,-*] ++(1.5,0) coordinate(H)
	to [C=$X_{\text{H}}$,-*] ++(0,-3);
	\draw (H)
	to [short] ++(1.5,0)
	to [R=$R_{\text{H}}$] ++(0,-3)
	to [short] (O);

	\draw [dashed] (-1.125,-0.5) rectangle (3.5,4);
	\draw [dashed] (4,-0.5) rectangle (9,4);
	\draw [dashed] (9.5,-0.5) rectangle (13.5,4);

	\draw (1.125,4.25) node[]{Antenna};
	\draw (6.5,4.25) node[]{Modulator};
	\draw (11.5,4.25) node[]{Harvester \& Chip};

	\draw (3.375,2.675) to [short] (3.375,3.175) to [short] (3.25,3.175) to [short,i_=$Z_{\text{A}}$] (3.125,3.175);
	\draw (4.125,-0.125) to [short] (4.125,0.375) to [short] (4.25,0.375) to [short,i=$Z_m$] (4.375,0.375);
	\draw (9.625,-0.125) to [short] (9.625,0.375) to [short] (9.75,0.375) to [short,i=$Z_{\text{H}}$] (9.875,0.375);
\end{circuitikz}
				}
				\label{fg:equivalent_circuit}
			}
			\subfloat[Scatter Model]{
				\resizebox{0.28\linewidth}{!}{
					\begin{circuitikz}[transform shape]
	\tikzstyle{every node}=[font=\large]
	\ctikzset{multipoles/rotary/arrow=both}
	\draw (0,0) node[bareantenna](bareantenna){};
	\draw (bareantenna.west) ++(-1,0) node[waves](WI){};
	\draw (WI.north east) ++(0.25,0) node{$\vec{E}_{\text{I}}$};
	\draw (WI.south east) ++(0.25,0) node{$\vec{H}_{\text{I}}$};
	\draw (bareantenna.east) ++(0.5,0) node[waves](WR){};
	\draw (WR.north east) ++(0.35,0) node{$\vec{E}_m$};
	\draw (WR.south east) ++(0.35,0) node{$\vec{H}_m$};
	\draw (bareantenna) to ++(0,-1.1) to [generic,l=$Z_{\text{A}}$,/tikz/circuitikz/bipoles/length=0.75cm] ++(2.3,0) node[rotary switch=4 in 90 wiper 30,anchor=ext center,rotate=0](SW){};
	\draw (SW.cout 1) to ++(0.625,+0.5) to [generic,/tikz/circuitikz/bipoles/length=0.75cm] ++(2.2,0) coordinate(E1);
	\draw (SW.cout 2) node[above right]{$m$} to [generic,l=$Z_m$,/tikz/circuitikz/bipoles/length=0.75cm] ++(2.2,0) coordinate(E2);
	\draw (SW.cout 3) to [generic,/tikz/circuitikz/bipoles/length=0.75cm] ++(2.2,0) coordinate(E3);
	\draw (SW.cout 4) to ++(0.625,-0.5) to [generic,/tikz/circuitikz/bipoles/length=0.75cm] ++(2.2,0) coordinate(E4);
	\draw (E1) to[short,-*] (E2);
	\draw (E2) to[short,-*] (E3);
	\draw (E3) to[short] (E4);
	\draw (E2) to[short] ++(0.2,0) node[ground,rotate=90]{};
\end{circuitikz}
				}
				\label{fg:scatter_model}
			}
			\caption{
			Block diagram, equivalent circuit, and scatter model of a RIScatter node.
			The solid and dashed vectors represent signal and energy flows.
			The scatter antenna behaves as a constant power source, where the voltage $V_0$ and current $I_0$ are introduced by incident electric field $\vec{E}_{\text{I}}$ and magnetic field $\vec{H}_{\text{I}}$ \cite{Huang2021}.
			}
			\label{fg:riscatter_node}
		\end{figure*}
	\end{subsection}

	\begin{subsection}{System Model}
		\label{sc:system_model}
		\begin{figure}[!t]
			\centering
			\def\svgwidth{0.7\columnwidth}
			\footnotesize{
				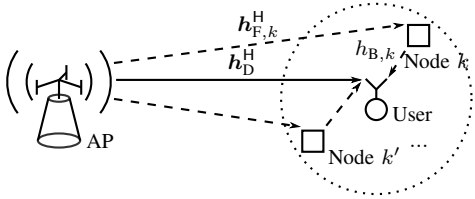
			}
			\caption{A single-user multi-node RIScatter network.}
			\label{fg:riscatter_network}
		\end{figure}
		As shown in Fig.~\ref{fg:riscatter_network}, we consider a RIScatter network where a $Q$-antenna \gls{ap} serves a single-antenna user and $K$ nearby dispersed or co-located RIScatter nodes.
		Without loss of generality, we assume all nodes have $M$ available reflection states.
		In the primary point-to-point system, the \gls{ap} transmits information to the user over a multipath channel\footnote{It is assumed the primary symbol duration is much longer than multipath delay spread (i.e., no inter-symbol interference).} enhanced by RIScatter nodes.
		In the backscatter multiple access system, the \gls{ap} acts as a carrier emitter, the RIScatter nodes modulate over the scattered signal, and the user jointly decodes all nodes.\footnote{It is assumed the signal going through two or more RIScatter nodes is too weak to be received by the user.}
		For simplicity, we consider a quasi-static block fading model and focus on a specific channel block where the \gls{csi} remains constant.
		Denote the \gls{ap}-user direct channel as $\boldsymbol{h}_{\text{D}}^\mathsf{H} \in \mathbb{C}^{1 \times Q}$, the \gls{ap}-node $k \in \mathcal{K} \triangleq \{1,\ldots,K\}$ forward channel as $\boldsymbol{h}_{\text{F},k}^\mathsf{H} \in \mathbb{C}^{1 \times Q}$, the node $k$-user backward channel as $h_{\text{B},k}$, and the cascaded \gls{ap}-node $k$-user channel as $\boldsymbol{h}_{\text{C},k}^\mathsf{H} \triangleq h_{\text{B},k} \boldsymbol{h}_{\text{F},k}^\mathsf{H} \in \mathbb{C}^{1 \times Q}$.
		We assume the direct and cascaded \gls{csi} are available at the \gls{ap} and user\footnote{The cascaded \gls{csi} can be estimated by sequential \cite{Bharadia2015,Yang2015b,Guo2019g} or parallel \cite{Jin2021a} approaches for dispersed nodes, or group-based \cite{Zheng2019} or hierarchical \cite{You2019} approaches for co-located nodes. The impact of channel estimation error will be investigated in Section~\ref{sc:simulation_results}.}.

		Let $\alpha_k \in \mathbb{I}$ be the amplitude scattering ratio of node $k$, $x_k \in \mathcal{X} \triangleq \{c_1,\ldots,c_M\}$ be the (coded) backscatter symbol of node $k$, and $x_{\mathcal{K}} \triangleq (x_1,\ldots,x_K)$ be the backscatter symbol tuple of all nodes.
		Due to double modulation, the composite channel is a function of backscatter symbol tuple\footnote{\eqref{eq:equivalent_channel_bc} and \eqref{eq:equivalent_channel_ris} are often used in \gls{bc} and \gls{ris} literature, respectively.}
		\begin{subequations}
			\label{eq:equivalent_channel}
			\begin{align}
				\boldsymbol{h}^\mathsf{H}(x_{\mathcal{K}})
				 & \triangleq \boldsymbol{h}_{\text{D}}^\mathsf{H} + \sum_{k} \alpha_k \boldsymbol{h}_{\text{C},k}^\mathsf{H} x_k \label{eq:equivalent_channel_bc}                    \\
				 & = \boldsymbol{h}_{\text{D}}^\mathsf{H} + \boldsymbol{x}^\mathsf{H} \mathrm{diag}(\boldsymbol{\alpha}) \boldsymbol{H}_{\text{C}}, \label{eq:equivalent_channel_ris}
			\end{align}
		\end{subequations}
		where $\boldsymbol{\alpha} \triangleq [\alpha_1,\ldots,\alpha_K]^\mathsf{T} \in \mathbb{I}^{K}$, $\boldsymbol{x} \triangleq [x_1,\ldots,x_K]^\mathsf{H} \in \mathcal{X}^{K}$, and $\boldsymbol{H}_{\text{C}} \triangleq [\boldsymbol{h}_{\text{C},1},\ldots,\boldsymbol{h}_{\text{C},K}]^\mathsf{H} \in \mathbb{C}^{K \times Q}$.
		Without loss of generality, we assume the spreading factor $N$ is a positive integer.
		Within one backscatter block, the signal received by the user at primary block $n \in \mathcal{N} \triangleq \{1,\ldots,N\}$ is
		\begin{equation}
			y[n] = \boldsymbol{h}^\mathsf{H}(x_{\mathcal{K}}) \boldsymbol{w} s[n] + v[n],
			\label{eq:receive_signal}
		\end{equation}
		where $\boldsymbol{w} \in \mathbb{C}^{Q}$ is the active beamformer satisfying $\lVert \boldsymbol{w} \rVert^2 \le P$, $P$ is the maximum average transmit power, $s \sim \mathcal{CN}(0,1)$ is the primary symbol, and $v \sim \mathcal{CN}(0,\sigma_v^2)$ is the \gls{awgn} with variance $\sigma_v^2$.

		Let $m_k \in \mathcal{M} \triangleq \{1,\ldots,M\}$ be the reflection state index of node $k$, and $m_{\mathcal{K}} \triangleq (m_1,\ldots,m_K)$ be the state index tuple of all nodes.
		The backscatter symbol $x_k$ (resp. symbol tuple $x_{\mathcal{K}}$) is a random variable that takes value $x_{m_k}$ (resp. value tuple $x_{m_{\mathcal{K}}}$) with probability $p(x_{m_k})$ (resp. $p(x_{m_{\mathcal{K}}})$).
		\begin{remark}
			Dispersed RIScatter nodes encode independently such that
			\begin{equation}
				p(x_{m_{\mathcal{K}}}) = \prod_k p(x_{m_k}).
				\label{eq:equivalent_distribution}
			\end{equation}
			When multiple nodes are co-located, they can jointly encode by designing the joint probability $p(x_{m_{\mathcal{K}}})$ directly.
			\label{re:independent_encoding}
		\end{remark}

		Let $z=\sum_{n} \bigl\lvert y[n] \bigr\rvert^2$ be the receive energy per backscatter block.
		When $x_{m_\mathcal{K}}$ is transmitted, the receive signal $y$ follows \gls{cscg} distribution $\mathcal{CN}(0,\sigma_{m_{\mathcal{K}}}^2)$ with variance
		\begin{equation}
			\sigma_{m_{\mathcal{K}}}^2 = \lvert \boldsymbol{h}^\mathsf{H}(x_{m_{\mathcal{K}}}) \boldsymbol{w} \rvert^2 + \sigma_v^2,
			\label{eq:receive_variance}
		\end{equation}
		and $z$ follows Gamma distribution with conditional \gls{pdf}
		\begin{equation}
			f(z|x_{m_{\mathcal{K}}}) = \frac{z^{N-1} \exp(-z/\sigma_{m_{\mathcal{K}}}^2)}{\sigma_{m_{\mathcal{K}}}^{2N} (N-1)!}.
			\label{eq:energy_distribution}
		\end{equation}

		\begin{remark}
			We have assumed Gaussian codebook for the primary source and finite support for the backscatter nodes, since they are relatively practical and widely adopted in relevant literatures \cite{Qian2019b,Xu2020b,Zhang2021d,Hua2022,Hu2021a,Qian2019}.
			The proposed framework is extendable to non-Gaussian primary source, and the conditional \gls{pdf} \eqref{eq:energy_distribution} can be approximated using \gls{clt} for large $N$.
			\label{re:non_gaussian}
		\end{remark}

		The user first jointly decodes the backscatter message of all nodes using a low-complexity energy detector.\footnote{The reliability of the energy detector is improved by the adaptive input distribution and thresholding design. With high-order modulation or large number of scatter nodes, the reliability can be further enhanced by increasing the spreading factor or using error correction codes with low code-rate. In practice, users can decode backscatter nodes ranging from a few to hundreds of meters in the presence of noise and interference, and the backscatter throughput can reach few Kbps to tens of Mbps \cite{Wu2022e}.\label{fn:energy_detector}}
		The energy detector formulates a \gls{dmtc} of size $M^K \times M^K$.

		\begin{remark}
			The capacity-achieving decision region design for \gls{dmtc} with non-binary inputs in arbitrary distribution remains an open issue.
			It was proved deterministic detectors can be rate-optimal, but non-convex regions (consist of non-adjacent partitions) are generally required and the optimal number of thresholds is unknown \cite{Nguyen2018,Nguyen2021}.
			Next, we restrict the energy detector to convex deterministic decision regions and consider sequential threshold design.
		\end{remark}

		Let $L = M^K$ be the number of decision regions.
		Sort $\{\sigma_{m_{\mathcal{K}}}^2\}$ in ascending order and denote the result sequence as $\sigma_1^2,\ldots,\sigma_L^2$.
		With sequential thresholding, the decision region of backscatter symbol tuple $l \in \mathcal{L} \triangleq \{1,\ldots,L\}$ is\footnote{$m_{\mathcal{K}}$ and $l$ are one-to-one mapped. Both notations are used interchangeably in the following context.}
		\begin{equation}
			\mathcal{R}_{l} \triangleq [t_{l-1},t_l), \quad 0 \le t_{l-1} \le t_l,
		\end{equation}
		where $t_l$ is the decision threshold between hypotheses $x_{l}$ and $x_{l+1}$.
		An example is shown in Fig.~\ref{fg:energy_distribution}.

		\begin{figure}[!t]
			\centering
			\includegraphics[width=0.8\columnwidth]{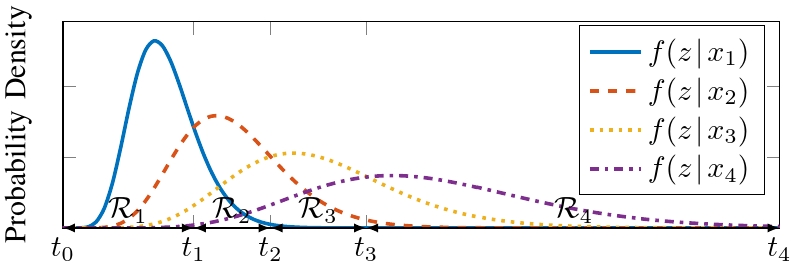}
			\caption{
				\gls{pdf} of the receive energy per backscatter block conditioned on different reflection state.
			}
			\label{fg:energy_distribution}
		\end{figure}
		When the threshold vector $\boldsymbol{t} \triangleq [t_0,\ldots,t_L]^\mathsf{T} \in \mathbb{R}_+^{L+1}$ is given, we can formulate a \gls{dmmac} with transition probability from input $x_{m_{\mathcal{K}}}$ to output $\hat{x}_{m_{\mathcal{K}}'}$ as
		\begin{equation}
			q(\hat{x}_{m_{\mathcal{K}}'}|x_{m_{\mathcal{K}}}) = \int_{\mathcal{R}_{m_{\mathcal{K}}'}} f(z|x_{m_{\mathcal{K}}}) \, d z.
			\label{eq:dmmac}
		\end{equation}
		The backscatter mutual information is
		\begin{equation}
			I_{\text{B}}(x_{\mathcal{K}};\hat{x}_{\mathcal{K}}) = \sum_{m_{\mathcal{K}}} p(x_{m_{\mathcal{K}}}) I_{\text{B}}(x_{m_{\mathcal{K}}};\hat{x}_{\mathcal{K}}),
			\label{eq:backscatter_mutual_information}
		\end{equation}
		where $I_{\text{B}}(x_{m_{\mathcal{K}}};\hat{x}_{\mathcal{K}})$ is the backscatter information function
		\begin{equation}
			I_{\text{B}}(x_{m_{\mathcal{K}}};\hat{x}_{\mathcal{K}}) \triangleq \sum_{m_{\mathcal{K}}'} q(\hat{x}_{m_{\mathcal{K}}'}|x_{m_{\mathcal{K}}}) \log \frac{q(\hat{x}_{m_{\mathcal{K}}'}|x_{m_{\mathcal{K}}})}{p(\hat{x}_{m_{\mathcal{K}}'})}.
			\label{eq:backscatter_information_function}
		\end{equation}

		Once the backscatter information is successfully decoded, the user re-encodes to recover the reflection pattern, constructs the composite channel by \eqref{eq:equivalent_channel}, then coherently decodes the primary link.
		The primary mutual information is
		\begin{equation}
			I_{\text{P}}(s;y|x_{\mathcal{K}}) = \sum_{m_{\mathcal{K}}} p(x_{m_{\mathcal{K}}}) I_{\text{P}}(s;y|x_{m_{\mathcal{K}}}),
			\label{eq:primary_mutual_information}
		\end{equation}
		where $I_{\text{P}}(s;y|x_{m_{\mathcal{K}}})$ is the primary information function
		\begin{equation}
			I_{\text{P}}(s;y|x_{m_{\mathcal{K}}}) \triangleq \log \Bigl(1 + \frac{\lvert \boldsymbol{h}^\mathsf{H}(x_{m_{\mathcal{K}}}) \boldsymbol{w} \rvert^2}{\sigma_v^2}\Bigr).
			\label{eq:primary_information_function}
		\end{equation}
	\end{subsection}
\end{section}

\begin{section}{Rate-Region Characterization}
	With a slight abuse of notation, we define the weighed sum mutual information and information function as
	\begin{align}
		I(x_{\mathcal{K}})
		 & \triangleq \rho I_{\text{P}}(s;y|x_{\mathcal{K}}) + (1 - \rho) I_{\text{B}}(x_{\mathcal{K}};\hat{x}_{\mathcal{K}}),\label{eq:weighted_sum_mutual_information}           \\
		I(x_{m_\mathcal{K}})
		 & \triangleq \rho I_{\text{P}}(s;y|x_{m_{\mathcal{K}}}) + (1 - \rho) I_{\text{B}}(x_{m_{\mathcal{K}}};\hat{x}_{\mathcal{K}}),\label{eq:weighted_sum_information_function}
	\end{align}
	where $\rho \in \mathbb{I}$ is the \gls{qos}.
	To obtain the achievable primary-(total-)backscatter rate region, we consider the weighted sum mutual information maximization problem with independent encoding at all nodes\footnote{Joint encoding over multiple nodes can be viewed as its special case with an augmented backscatter source.\label{fn:joint_encoding}}
	\begin{maxi!}
		{\scriptstyle{\{\boldsymbol{p}_k\}_{k \in \mathcal{K}},\boldsymbol{w},\boldsymbol{t}}}{I(x_{\mathcal{K}})}{\label{op:weighted_sum_rate}}{\label{ob:weighted_sum_rate}}
		\addConstraint{\boldsymbol{1}^\mathsf{T} \boldsymbol{p}_k}{=1,}{\quad \forall k}{\label{co:sum_probability}}
		\addConstraint{\boldsymbol{p}_k}{\ge \boldsymbol{0},}{\quad \forall k}{\label{co:nonnegative_probability}}
		\addConstraint{\lVert \boldsymbol{w} \rVert^2}{\le P}{\label{co:transmit_power}}
		\addConstraint{t_{l-1}}{\le t_l,}{\quad \forall l}{\label{co:sequential_threshold}}
		\addConstraint{\boldsymbol{t}}{\ge \boldsymbol{0},}{\label{co:nonnegative_threshold}}
	\end{maxi!}
	where $\boldsymbol{p}_k = [p(x_{1_k}), \ldots, p(x_{M_k})]^\mathsf{T} \in \mathbb{I}^M$ is the input distribution of node $k$.
	Problem \eqref{op:weighted_sum_rate} generalizes \gls{bc} by allowing \gls{csi}- and \gls{qos}-adaptive input distribution and decision region design.
	On the other hand, it also relaxes the feasible domain of discrete \gls{ris} phase shift selection problems from the vertices of $M$-dimensional probability simplex to the simplex itself.
	The original problem is highly non-convex and we propose a \gls{bcd} algorithm that iteratively updates $\{\boldsymbol{p}_k\}_{k \in \mathcal{K}}$, $\boldsymbol{w}$ and $\boldsymbol{t}$.

	\begin{subsection}{Input Distribution}
		\label{sc:input_distribution}
		For any given $\boldsymbol{w}$ and $\boldsymbol{t}$, we can formulate a \gls{dmmac} by \eqref{eq:dmmac} and simplify \eqref{op:weighted_sum_rate} to
		\begin{maxi!}
			{\scriptstyle{\{\boldsymbol{p}_k\}_{k \in \mathcal{K}}}}{I(x_{\mathcal{K}})}{\label{op:input_distribution}}{}
			\addConstraint{\eqref{co:sum_probability},\eqref{co:nonnegative_probability},}
		\end{maxi!}
		which involves the product term \eqref{eq:equivalent_distribution} and is generally non-convex (unless $K=1$).
		Following \cite{Rezaeian2004}, we first recast the \gls{kkt} conditions to their equivalent forms, then propose a numerical solution that guarantees those conditions on the converging point of a sequence.
		\begin{proposition}
			The \gls{kkt} optimality conditions for problem \eqref{op:input_distribution} are equivalent to, $\forall k,m_k$,
			\begin{subequations}
				\label{eq:input_kkt_condition}
				\begin{alignat}{2}
					I_k^\star(x_{m_k}) & = I^\star(x_{\mathcal{K}}), \quad   &  & \text{if} \ p^\star(x_{m_k}) > 0,\label{eq:probable_states} \\
					I_k^\star(x_{m_k}) & \le I^\star(x_{\mathcal{K}}), \quad &  & \text{if} \ p^\star(x_{m_k}) = 0,\label{eq:dropped_states}
				\end{alignat}
			\end{subequations}
			where $I_k(x_{m_k})$ is the weighted sum marginal information
			\begin{align}
				I_k(x_{m_k})
				 & \triangleq \sum_{m_{\mathcal{K} \setminus \{k\}}} p(x_{m_{\mathcal{K} \setminus \{k\}}}) I(x_{m_\mathcal{K}}).
				\label{eq:weighted_sum_marginal_information}
			\end{align}
			\label{pr:input_kkt_condition}
		\end{proposition}

		\begin{proof}
			Please refer to Appendix \ref{ap:input_kkt_condition}.
			\label{pf:input_kkt_condition}
		\end{proof}

		For each RIScatter node, \eqref{eq:probable_states} suggests each probable state should produce the same marginal information (averaged over all states of other nodes), while \eqref{eq:dropped_states} suggests any state with potentially less marginal information should not be used.
		\begin{proposition}
			For any strictly positive initializer $\{\boldsymbol{p}_k^{(0)}\}_{k \in \mathcal{K}}$, the \gls{kkt} input probability of node $k$ at state $m_k$ is given by the converging point of the sequence
			\begin{equation}
				p^{(r+1)}(x_{m_k}) = \frac{p^{(r)}(x_{m_k}) \exp \Bigl( \frac{\rho}{1 - \rho} I_k^{(r)}(x_{m_k}) \Bigr)}{\sum_{m_k'} p^{(r)}(x_{m_k'}) \exp \Bigl( \frac{\rho}{1 - \rho} I_k^{(r)}(x_{m_k'}) \Bigr)},
				\label{eq:input_kkt_solution}
			\end{equation}
			where $r$ is the iteration index.
			\label{pr:input_kkt_solution}
		\end{proposition}
		\begin{proof}
			Please refer to Appendix \ref{ap:input_kkt_solution}.
			\label{pf:input_kkt_solution}
		\end{proof}

		At iteration $r+1$, the input distribution of node $k$ is updated over $\bigl\{\{\boldsymbol{p}_q^{(r+1)}\}_{q=1}^{k-1},\{\boldsymbol{p}_q^{(r)}\}_{q=k}^{K}\bigr\}$.
		The \gls{kkt} input distribution design is summarized in Algorithm \ref{al:input_distribution}.

		\setalgorithmcaptionfont{\small}
		\begin{algorithm}[!t]
			\small
			\caption{Input Distribution Evaluation by a Sequence}
			\label{al:input_distribution}
			\begin{algorithmic}[1]
				\Require $K$, $N$, $\boldsymbol{h}_{\text{D}}^\mathsf{H}$, $\boldsymbol{H}_{\text{C}}$, $\boldsymbol{\alpha}$, $\mathcal{X}$, $\sigma_v^2$, $\rho$, $\boldsymbol{w}$, $\boldsymbol{t}$, $\epsilon$
				\Ensure $\{\boldsymbol{p}_k^\star\}_{k \in \mathcal{K}}$
				\State Set $\boldsymbol{h}^\mathsf{H}(x_{m_{\mathcal{K}}})$, $\forall m_{\mathcal{K}}$ by \eqref{eq:equivalent_channel}
				\State \phantom{Set} $\sigma^2_{m_{\mathcal{K}}}$, $\forall m_{\mathcal{K}}$ by \eqref{eq:receive_variance}
				\State \phantom{Set} $f(z|x_{m_{\mathcal{K}}})$, $\forall m_{\mathcal{K}}$ by \eqref{eq:energy_distribution}
				\State \phantom{Set} $q(\hat{x}_{m_{\mathcal{K}}'}|x_{m_{\mathcal{K}}})$, $\forall m_{\mathcal{K}}, m_{\mathcal{K}}'$ by \eqref{eq:dmmac}
				\State Initialize $r \gets 0$
				\State \phantom{Initialize} $\boldsymbol{p}_k^{(0)} > \boldsymbol{0}$, $\forall k$
				\State Get $p^{(r)}(x_{m_{\mathcal{K}}})$, $\forall m_{\mathcal{K}}$ by \eqref{eq:equivalent_distribution} \label{st:input_distribution_begin}
				\State \phantom{Get} $I^{(r)}(x_{m_{\mathcal{K}}})$, $\forall m_{\mathcal{K}}$ by \eqref{eq:backscatter_information_function}, \eqref{eq:primary_information_function}, \eqref{eq:weighted_sum_information_function}
				\State \phantom{Get} $I^{(r)}_k(x_{m_k})$, $\forall k,m_k$ by \eqref{eq:weighted_sum_marginal_information}
				\State \phantom{Get} $I^{(r)}(x_{\mathcal{K}})$ by \eqref{eq:backscatter_mutual_information}, \eqref{eq:primary_mutual_information}, \eqref{eq:weighted_sum_mutual_information} \label{st:input_distribution_end}
				\Repeat
					\State Update $r \gets r+1$
					\State \phantom{Update} $\boldsymbol{p}_k^{(r)}$, $\forall k$ by \eqref{eq:input_kkt_solution}
					\State Redo step \ref{st:input_distribution_begin}--\ref{st:input_distribution_end}
				\Until $I^{(r)}(x_{\mathcal{K}}) - I^{(r-1)}(x_{\mathcal{K}}) \le \epsilon$
			\end{algorithmic}
		\end{algorithm}

		\begin{remark}
			The insufficiency of the \gls{kkt} conditions for problem \eqref{op:input_distribution} implies that the proposed method may not converge to the global-optimal solution.
			However, simulation results in Section~\ref{sc:simulation_results} will show that the average performance gap is indistinguishable at a moderate $K$.
			\label{re:input_kkt_distribution}
		\end{remark}
	\end{subsection}

	\begin{subsection}{Active Beamforming}
		For any given $\{\boldsymbol{p}_k\}_{k \in \mathcal{K}}$ and $\boldsymbol{t}$, problem \eqref{op:weighted_sum_rate} reduces to
		\begin{maxi!}
			{\scriptstyle{\boldsymbol{w}}}{I(x_{\mathcal{K}})}{\label{op:active_beamforming}}{\label{ob:active_beamforming}}
			\addConstraint{\eqref{co:transmit_power},}
		\end{maxi!}
		which is still non-convex due to the integration and entropy terms.
		Note the \gls{dmmac} $q(x_l|x_{m_{\mathcal{K}}})$ depends on the variance of accumulated receive energy $\sigma_{m_{\mathcal{K}}}^2$, which is a function of $\boldsymbol{w}$.
		Plugging \eqref{eq:energy_distribution} into \eqref{eq:dmmac}, we have
		\begin{align}
			q(x_l|x_{m_{\mathcal{K}}})
			 & = \frac{\int_{{t_{l-1}}/{\sigma_{m_{\mathcal{K}}}^2}}^{{t_l}/{\sigma_{m_{\mathcal{K}}}^2}} z^{N-1} \exp(-z) \, d z}{(N-1)!} \\
			 & = Q\Bigl(N,\frac{t_{l-1}}{\sigma_{m_{\mathcal{K}}}^2},\frac{t_l}{\sigma_{m_{\mathcal{K}}}^2}\Bigr),
			\label{eq:dmmac_gamma}
		\end{align}
		where $Q(N, b_1, b_2) \triangleq \int_{b_1}^{b_2} z^{N-1} \exp(-z) / (N - 1)! \, d z$ is the regularized incomplete Gamma function.
		Its series expansion is given by \cite[Theorem 3]{Jameson2016}
		\begin{align}
			Q\Bigl(N,\frac{t_{l-1}}{\sigma_{m_{\mathcal{K}}}^2},\frac{t_l}{\sigma_{m_{\mathcal{K}}}^2}\Bigr)
			 & = \exp \Bigl(-\frac{t_{l-1}}{\sigma_{m_{\mathcal{K}}}^2}\Bigr) \sum_{n=0}^{N-1} \frac{\bigl(\frac{t_{l-1}}{\sigma_{m_{\mathcal{K}}}^2}\bigr)^n}{n!} \nonumber \\
			 & \quad - \exp \Bigl(-\frac{t_l}{\sigma_{m_{\mathcal{K}}}^2}\Bigr) \sum_{n=0}^{N-1} \frac{\bigl(\frac{t_l}{\sigma_{m_{\mathcal{K}}}^2}\bigr)^n}{n!},
			\label{eq:regularized_incomplete_gamma}
		\end{align}
		whose gradient with respect to $\boldsymbol{w}^*$ is
		\begin{align}
			\nabla_{\boldsymbol{w}^*} Q\Bigl(N,\frac{t_{l-1}}{\sigma_{m_{\mathcal{K}}}^2},\frac{t_l}{\sigma_{m_{\mathcal{K}}}^2}\Bigr)
			 & = \frac{\boldsymbol{h}(x_{m_{\mathcal{K}}})\boldsymbol{h}^\mathsf{H}(x_{m_{\mathcal{K}}})\boldsymbol{w}}{(\sigma_{m_{\mathcal{K}}}^2)^2} \nonumber \\
			 & \quad \times \Bigl(g_{m_{\mathcal{K}}}(t_l) - g_{m_{\mathcal{K}}}(t_{l-1})\Bigr),
			\label{eq:regularized_incomplete_gamma_gradient}
		\end{align}
		where
		\begin{equation}
			g_{m_{\mathcal{K}}}(t_l) = t_l\exp\Bigl(-\frac{t_l}{\sigma_{m_{\mathcal{K}}}^2}\Bigr)\Bigl(-1+\sum_{n=1}^{N-1} \frac{\bigl(n - \frac{t_l}{\sigma_{m_{\mathcal{K}}}^2}\bigr) \bigl(\frac{t_l}{\sigma_{m_{\mathcal{K}}}^2}\bigr)^{n-1}}{n!}\Bigr).
			\label{eq:regularized_incomplete_gamma_gradient_component}
		\end{equation}
		On top of \eqref{eq:regularized_incomplete_gamma} and \eqref{eq:regularized_incomplete_gamma_gradient}, we rewrite $I(x_{\mathcal{K}})$ and $\nabla_{\boldsymbol{w}^*} I(x_{\mathcal{K}})$ as \eqref{eq:weighted_sum_mutual_information_explicit} and \eqref{eq:weighted_sum_mutual_information_gradient} at the end of page \pageref{eq:weighted_sum_mutual_information_explicit}, respectively.
		\begin{figure*}[!b]
			\hrule
			\begin{equation}
				I(x_{\mathcal{K}})=\sum_{m_{\mathcal{K}}}p(x_{m_{\mathcal{K}}})\Biggl(\rho\log\Bigl(1+\frac{\lvert\boldsymbol{h}^\mathsf{H}(x_{m_{\mathcal{K}}})\boldsymbol{w}\rvert^2}{\sigma_v^2}\Bigr)+(1-\rho)\sum_l Q\Bigl(N,\frac{t_{l-1}}{\sigma_{m_{\mathcal{K}}}^2},\frac{t_l}{\sigma_{m_{\mathcal{K}}}^2}\Bigr) \log \frac{Q\Bigl(N,\frac{t_{l-1}}{\sigma_{m_{\mathcal{K}}}^2},\frac{t_l}{\sigma_{m_{\mathcal{K}}}^2}\Bigr)}{\sum_{m_{\mathcal{K}}'} p(x_{m_{\mathcal{K}}'}) Q\Bigl(N,\frac{t_{l-1}}{\sigma_{m_{\mathcal{K}}'}^2},\frac{t_l}{\sigma_{m_{\mathcal{K}}'}^2}\Bigr)}\Biggr)
				\label{eq:weighted_sum_mutual_information_explicit}
			\end{equation}
			\begin{align}
				\nabla_{\boldsymbol{w}^*} I(x_{\mathcal{K}})
				 & = \sum_{m_{\mathcal{K}}}p(x_{m_{\mathcal{K}}})\Biggl(\rho\frac{\boldsymbol{h}(x_{m_{\mathcal{K}}})\boldsymbol{h}^\mathsf{H}(x_{m_{\mathcal{K}}})\boldsymbol{w}}{\sigma_{m_{\mathcal{K}}}^2}+(1-\rho)\sum_l\biggl(\log\frac{Q\Bigl(N,\frac{t_{l-1}}{\sigma_{m_{\mathcal{K}}}^2},\frac{t_l}{\sigma_{m_{\mathcal{K}}}^2}\Bigr)}{\sum_{m_{\mathcal{K}}'}p(x_{m_{\mathcal{K}}'})Q\Bigl(N,\frac{t_{l-1}}{\sigma_{m_{\mathcal{K}}'}^2},\frac{t_l}{\sigma_{m_{\mathcal{K}}'}^2}\Bigr)}+1\biggr)\nonumber                                                                                   \\
				 & \qquad \times \nabla_{\boldsymbol{w}^*} Q\Bigl(N,\frac{t_{l-1}}{\sigma_{m_{\mathcal{K}}}^2},\frac{t_l}{\sigma_{m_{\mathcal{K}}}^2}\Bigr)-\frac{Q\Bigl(N,\frac{t_{l-1}}{\sigma_{m_{\mathcal{K}}}^2},\frac{t_l}{\sigma_{m_{\mathcal{K}}}^2}\Bigr)\sum_{m_{\mathcal{K}}'}p(x_{m_{\mathcal{K}}'})\nabla_{\boldsymbol{w}^*}Q\Bigl(N,\frac{t_{l-1}}{\sigma_{m_{\mathcal{K}}'}^2},\frac{t_l}{\sigma_{m_{\mathcal{K}}'}^2}\Bigr)}{\sum_{m_{\mathcal{K}}'}p(x_{m_{\mathcal{K}}'})Q\Bigl(N,\frac{t_{l-1}}{\sigma_{m_{\mathcal{K}}'}^2},\frac{t_l}{\sigma_{m_{\mathcal{K}}'}^2}\Bigr)}\Biggr)
				\label{eq:weighted_sum_mutual_information_gradient}
			\end{align}
		\end{figure*}
		Problem \eqref{op:active_beamforming} can thus be solved by the \gls{pga} method.
		At iteration $r+1$, the unregulated active beamformer is updated by
		\begin{equation}
			\bar{\boldsymbol{w}}^{(r+1)} = \boldsymbol{w}^{(r)}+\gamma\nabla_{\boldsymbol{w}^*} I^{(r)}(x_{\mathcal{K}}),
			\label{eq:beamforming_gradient_ascent}
		\end{equation}
		where $\gamma$ is the step size (refinable by backtracking line search \cite[Section 9.2]{Boyd2004}).
		Then, $\bar{\boldsymbol{w}}$ is projected onto the feasible domain \eqref{co:transmit_power} to retrieve the active beamformer
		\begin{equation}
			\boldsymbol{w} = \frac{\sqrt{P} \bar{\boldsymbol{w}}}{\max\bigl(\sqrt{P},\lVert\bar{\boldsymbol{w}}\rVert\bigr)}.
			\label{eq:beamforming_projection}
		\end{equation}

		The \gls{pga} active beamforming design is summarized in Algorithm \ref{al:active_beamforming}.
		\setalgorithmcaptionfont{\small}
		\begin{algorithm}[!t]
			\small
			\caption{Active Beamforming Optimization by \gls{pga}}
			\label{al:active_beamforming}
			\begin{algorithmic}[1]
				\Require $Q$, $N$, $\boldsymbol{h}_{\text{D}}^\mathsf{H}$, $\boldsymbol{H}_{\text{C}}$, $\boldsymbol{\alpha}$, $\mathcal{X}$, $P$, $\sigma_v^2$, $\rho$, $\{\boldsymbol{p}_k\}_{k \in \mathcal{K}}$, $\boldsymbol{t}$, $\alpha$, $\beta$, $\gamma$, $\epsilon$
				\Ensure $\boldsymbol{w}^\star$
				\State Set $\boldsymbol{h}^\mathsf{H}(x_{m_{\mathcal{K}}})$, $\forall m_{\mathcal{K}}$ by \eqref{eq:equivalent_channel}
				\State \phantom{Set} $p(x_{m_{\mathcal{K}}})$, $\forall m_{\mathcal{K}}$ by \eqref{eq:equivalent_distribution}
				\State Initialize $r \gets 0$
				\State \phantom{Initialize} $\boldsymbol{w}^{(0)}$, $\lVert\boldsymbol{w}^{(0)}\rVert^2 \le P$
				\State Get $(\sigma_{m_{\mathcal{K}}}^{(r)})^2$, $\forall m_{\mathcal{K}}$ by \eqref{eq:receive_variance} \label{st:gradient_descent_begin}
				\State \phantom{Get} $Q^{(r)}\bigl(N,\frac{t_{l-1}}{\sigma_{m_{\mathcal{K}}}^2},\frac{t_l}{\sigma_{m_{\mathcal{K}}}^2}\bigr)$, $\forall m_{\mathcal{K}},l$ by \eqref{eq:regularized_incomplete_gamma}
				\State \phantom{Get} $I^{(r)}(x_{\mathcal{K}})$ by \eqref{eq:weighted_sum_mutual_information_explicit} \label{st:gradient_descent_end}
				\State \phantom{Get} $\nabla_{\boldsymbol{w}^*} Q^{(r)}\bigl(N,\frac{t_{l-1}}{\sigma_{m_{\mathcal{K}}}^2},\frac{t_l}{\sigma_{m_{\mathcal{K}}}^2}\bigr)$, $\forall m_{\mathcal{K}},l$ by \eqref{eq:regularized_incomplete_gamma_gradient} \label{st:gradient_update_start}
				\State \phantom{Get} $\nabla_{\boldsymbol{w}^*} I^{(r)}(x_{\mathcal{K}})$ by \eqref{eq:weighted_sum_mutual_information_gradient} \label{st:gradient_update_end}
				\Repeat
					\State Update $r \gets r+1$
					\State \phantom{Update} $\gamma^{(r)}\gets\gamma$
					\State \phantom{Update} $\bar{\boldsymbol{w}}^{(r)}$ by \eqref{eq:beamforming_gradient_ascent} \label{st:backtracking_line_search_begin}
					\State \phantom{Update} $\boldsymbol{w}^{(r)}$ by \eqref{eq:beamforming_projection}
					\State Redo step \ref{st:gradient_descent_begin}--\ref{st:gradient_descent_end} \label{st:backtracking_line_search_end}
					\While{$I^{(r)}(x_{\mathcal{K}})<I^{(r-1)}(x_{\mathcal{K}})+\alpha\gamma\lVert\nabla_{\boldsymbol{w}^*}I^{(r-1)}(x_{\mathcal{K}})\rVert^2$}
						\State Set $\gamma^{(r)}\gets\beta\gamma^{(r)}$
						\State Redo step \ref{st:backtracking_line_search_begin}--\ref{st:backtracking_line_search_end}
					\EndWhile
					\State Redo step \ref{st:gradient_update_start}, \ref{st:gradient_update_end}
				\Until $\lVert\boldsymbol{w}^{(r)}-\boldsymbol{w}^{(r-1)}\rVert \le \epsilon$
			\end{algorithmic}
		\end{algorithm}
	\end{subsection}

	\begin{subsection}{Decision Threshold}
		For any given $\{\boldsymbol{p}_k\}_{k \in \mathcal{K}}$ and $\boldsymbol{w}$, problem \eqref{op:weighted_sum_rate} reduces to
		\begin{maxi!}
			{\scriptstyle{\boldsymbol{t}}}{I(x_{\mathcal{K}})}{\label{op:decision_threshold}}{\label{ob:decision_threshold}}
			\addConstraint{\eqref{co:sequential_threshold},\eqref{co:nonnegative_threshold},}
		\end{maxi!}
		which is still non-convex since $\boldsymbol{t}$ appears on the limits of integration \eqref{eq:dmmac}.
		Instead of solving it directly, we constrain the feasible domain from continuous space $\mathbb{R}_+^{L+1}$ to discrete candidates (i.e., fine-grained energy levels) $\mathcal{T}^{L+1}$.
		As shown in Fig.~\ref{fg:discrete_outputs}, the decision regions are formulated by grouping adjacent energy bins.
		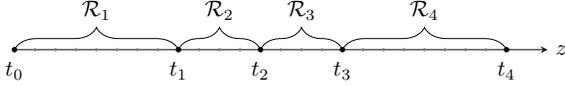
\begin{figure}[!t]
			\centering
			\resizebox{0.9\columnwidth}{!}{
				\begin{tikzpicture}
	\begin{axis}[width=10cm,height=5cm,xmin=0,xmax=6.5,ymin=-0.1,ymax=0.1,xlabel=$z$,axis x line=middle,axis y line=none,xtick={0,2,3,4,6},xticklabels={$t_0$,$t_1$,$t_2$,$t_3$,$t_4$},xtick style={draw=none},every axis x label/.style={at=(current axis.right of origin),anchor=west},domain=0:6]
		\addplot[mark=*,mark size=0.5pt,only marks,color=gray] {0};
		\addplot[mark=*,mark size=1pt,only marks,color=black] coordinates{(0,0) (2,0) (3,0) (4,0) (6,0)};
		\draw[decorate,decoration={brace,amplitude=10}] (0,0) -- (2,0) node[above,midway,yshift=10]{$\mathcal{R}_1$};
		\draw[decorate,decoration={brace,amplitude=10}] (2,0) -- (3,0) node[above,midway,yshift=10]{$\mathcal{R}_2$};
		\draw[decorate,decoration={brace,amplitude=10}] (3,0) -- (4,0) node[above,midway,yshift=10]{$\mathcal{R}_3$};
		\draw[decorate,decoration={brace,amplitude=10}] (4,0) -- (6,0) node[above,midway,yshift=10]{$\mathcal{R}_4$};
	\end{axis}
\end{tikzpicture}
			}
			\caption{The thresholds are chosen from fine-grained candidates instead of the continuous space. Each decision region consists of at least one bin.}
			\label{fg:discrete_outputs}
		\end{figure}

		\begin{remark}
			The design of the energy detector does not affect the primary achievable rate, since the composite channel \eqref{eq:equivalent_channel} can always be determined after backscatter decoding and re-encoding.
			This implies that any thresholding maximizing the total backscatter rate is optimal for problem \eqref{op:decision_threshold}.
			\label{re:backscatter_decision}
		\end{remark}

		\begin{remark}
			In terms of total backscatter rate, the nodes can be viewed as an augmented source, and problem \eqref{op:decision_threshold} becomes the rate-optimal quantizer design for \gls{dmtc}.
			\label{re:augmented_source}
		\end{remark}

		Thanks to Remark \ref{re:backscatter_decision} and \ref{re:augmented_source}, problem \eqref{op:decision_threshold} can be recast as
		\begin{maxi!}
			{\scriptstyle{\boldsymbol{t} \in \mathcal{T}^{L+1}}}{I_{\text{B}}(x_{\mathcal{K}};\hat{x}_{\mathcal{K}})}{\label{op:decision_threshold_discrete}}{\label{ob:decision_threshold_discrete}}
			\addConstraint{\eqref{co:sequential_threshold},}
		\end{maxi!}
		whose global optimal solution has been obtained in recent works.
		\cite{He2021} started from the quadrangle inequality and proposed a \gls{dp} method accelerated by the \gls{smawk} algorithm with computational complexity $\mathcal{O}\bigl(L^2(\mathrm{card}(\mathcal{T})-L)\bigr)$.
		On the other hand, \cite{Nguyen2020a} started from the optimality condition for three neighbor thresholds and presented a traverse-then-bisect algorithm with complexity $\mathcal{O}\bigl(\mathrm{card}(\mathcal{T})L\log(\mathrm{card}(\mathcal{T})L)\bigr)$.
		In Section \ref{sc:simulation_results}, both schemes will be compared with the \gls{ml} scheme \cite{Qian2019}
		\begin{equation}
			t_{l}^{\text{\gls{ml}}} = N \frac{\sigma_{l-1}^2 \sigma_{l}^2}{\sigma_{l-1}^2 - \sigma_{l}^2} \log \frac{\sigma_{l-1}^2}{\sigma_{l}^2}, \quad l \in \mathcal{L} \setminus \{L\},
			\label{eq:detection_threshold_ml}
		\end{equation}
		which is suboptimal for problem \eqref{op:decision_threshold} unless all nodes are with equiprobable inputs.
	\end{subsection}
\end{section}

\begin{section}{Simulation Results}
	\label{sc:simulation_results}
	In this section, we provide numerical results to evaluate the proposed algorithms.
	We assume the AP-user distance is \qty{10}{m} and at least one RIScatter nodes are randomly dropped in a disk centered at the user with radius $r$.
	The \gls{ap} is with maximum average transmit power $P=\qty{36}{dBm}$ and all nodes employs $M$-\gls{qam} with $\alpha=0.5$.
	For all channels involved, we consider a distance-dependent path loss model
	\begin{equation}
		L(d) = L_0 \biggl(\frac{d_0}{d}\biggr)^\gamma,
	\end{equation}
	together with a Rician fading model
	\begin{equation}
		\boldsymbol{H} = \sqrt{\frac{\kappa}{1+\kappa}} \bar{\boldsymbol{H}} + \sqrt{\frac{1}{1+\kappa}} \tilde{\boldsymbol{H}},
	\end{equation}
	where $d$ is the transmission distance, $L_0=-\qty{30}{dB}$ is the reference path loss at $d_0=\qty{1}{m}$, $\kappa$ is the Rician K-factor, $\bar{\boldsymbol{H}}$ is the deterministic line-of-sight component with unit-magnitude entries, and $\tilde{\boldsymbol{H}}$ is the Rayleigh fading component with standard \gls{iid} \gls{cscg} entries.
	We choose $\gamma_{\text{D}}=2.6$, $\gamma_{\text{F}}=2.4$, $\gamma_{\text{B}}=2$, and $\kappa_{\text{D}}=\kappa_{\text{F}}=\kappa_{\text{B}}=5$ for direct, forward and backward links.
	The finite decision threshold domain $\mathcal{T}$ is obtained by $b$-bit uniform discretization over the critical interval defined by the $1-\varepsilon$ confidence bounds of edge hypotheses (i.e., lower bound of $x_1$ and upper bound of $x_L$).
	We set $b=9$ and $\varepsilon=\num{e-3}$.
	All achievable rate regions are averaged over \num{e3} channel realizations.\footnote{The code is publicly available at \url{https://github.com/snowztail/riscatter/}.}

	\begin{subsection}{Evaluation of Proposed Algorithms}
		\begin{subsubsection}{Initialization}
			To characterize the achievable rate region, we progressively obtain all boundary points by successively increasing $\rho$ and solving problem \eqref{op:weighted_sum_rate}.
			For $\rho=0$ where the backscatter link is prioritized, we initialize Algorithm \ref{al:input_distribution} and \ref{al:active_beamforming} by uniform input distribution and \gls{mrt} towards the sum cascaded channel $\sum_{k} \boldsymbol{h}_{\text{C},k}^\mathsf{H}$, respectively.
			At the following points, both algorithms are initialized by the solutions at the previous point.
		\end{subsubsection}

		\begin{subsubsection}{Convergence}
			\label{sc:convergence}
			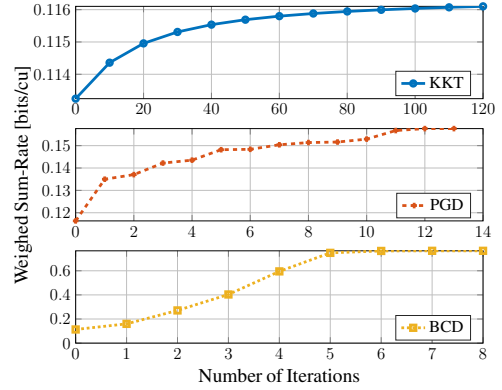
\begin{figure}[!t]
				\centering
				\resizebox{0.75\columnwidth}{!}{
%
%
\definecolor{mycolor1}{rgb}{0.00000,0.44706,0.74118}%
\definecolor{mycolor2}{rgb}{0.85098,0.32549,0.09804}%
\definecolor{mycolor3}{rgb}{0.92941,0.69412,0.12549}%
\begin{tikzpicture}

	\begin{axis}[%
			width=4.053in,
			height=0.919in,
			at={(0.68in,2.887in)},
			scale only axis,
			xmin=0,
			xmax=120,
			ymin=0.113244880454575,
			ymax=0.116102304831398,
			axis background/.style={fill=white},
			xmajorgrids,
			ymajorgrids,
			legend style={at={(0.97,0.03)}, anchor=south east, legend cell align=left, align=left, draw=white!15!black},
			title style={font=\Large},
			label style={font=\Large},
			ticklabel style={font=\large},
			legend style={font=\large},
			yticklabel=\pgfkeys{/pgf/number format/.cd,fixed,precision=3}\pgfmathprintnumber{\tick}
		]
		\addplot [color=mycolor1, line width=2.0pt, mark=o, mark options={solid, mycolor1}]
		table[row sep=crcr]{%
				0	0.113244880454575\\
				10	0.114358992707155\\
				20	0.114956255981364\\
				30	0.115310191711237\\
				40	0.115536251457935\\
				50	0.11568965371294\\
				60	0.115799285291346\\
				70	0.115881218617581\\
				80	0.115944814587431\\
				90	0.115995739004593\\
				100	0.116037546854275\\
				110	0.116072549598378\\
				120	0.116102304831398\\
			};
		\addlegendentry{KKT}

	\end{axis}

	\begin{axis}[%
		width=4.053in,
		height=0.919in,
		at={(0.68in,1.67in)},
		scale only axis,
		xmin=0,
		xmax=14,
		ymin=0.116354276985174,
		ymax=0.157617738080098,
		ylabel style={font=\color{white!15!black}},
		ylabel={Weighed Sum-Rate [bits/cu]},
		axis background/.style={fill=white},
		xmajorgrids,
		ymajorgrids,
		legend style={at={(0.97,0.03)}, anchor=south east, legend cell align=left, align=left, draw=white!15!black},
		title style={font=\Large},
		label style={font=\Large},
		ticklabel style={font=\large},
		legend style={font=\large},
		yticklabel=\pgfkeys{/pgf/number format/.cd,fixed,precision=3}\pgfmathprintnumber{\tick}
		]
		\addplot [color=mycolor2, dashed, line width=2.0pt, mark=+, mark options={solid, mycolor2}]
		table[row sep=crcr]{%
				0	0.116354276985174\\
				1	0.134986216301261\\
				2	0.137045463164113\\
				3	0.14221149883543\\
				4	0.143513293823474\\
				5	0.148182976472748\\
				6	0.148354041018766\\
				7	0.150388146565952\\
				8	0.151376635136393\\
				9	0.15160915733737\\
				10	0.15291935995385\\
				11	0.156743857286177\\
				12	0.157617738080098\\
				13	0.157617738080098\\
			};
		\addlegendentry{PGD}

	\end{axis}

	\begin{axis}[%
			width=4.053in,
			height=0.919in,
			at={(0.68in,0.453in)},
			scale only axis,
			xmin=0,
			xmax=8,
			xlabel style={font=\color{white!15!black}},
			xlabel={Number of Iterations},
			ymin=0,
			ymax=0.765555326497281,
			axis background/.style={fill=white},
			xmajorgrids,
			ymajorgrids,
			legend style={at={(0.97,0.03)}, anchor=south east, legend cell align=left, align=left, draw=white!15!black},
			title style={font=\Large},
			label style={font=\Large},
			ticklabel style={font=\large},
			legend style={font=\large},
			yticklabel=\pgfkeys{/pgf/number format/.cd,fixed,precision=3}\pgfmathprintnumber{\tick}
		]
		\addplot [color=mycolor3, dotted, line width=2.0pt, mark=square, mark options={solid, mycolor3}]
		table[row sep=crcr]{%
				0	0.113244880454575\\
				1	0.159820508450486\\
				2	0.271161604498627\\
				3	0.403558255259651\\
				4	0.594995557319089\\
				5	0.748644776092643\\
				6	0.764477923292286\\
				7	0.765555325740019\\
				8	0.765555326497281\\
			};
		\addlegendentry{BCD}

	\end{axis}
\end{tikzpicture}%
				}
				\caption{Typical convergence curves at $\rho=0$ for $Q=4$, $K=8$, $M=2$, $N=20$, $\sigma_v^2=\qty{-40}{dBm}$ and $r=\qty{2}{m}$.}
				\label{fg:wsr_convergence}
			\end{figure}
			The \gls{bcd} algorithm is convergent for problem \eqref{op:weighted_sum_rate} since the input distribution and active beamforming subproblems converge and the thresholding subproblem attains global optimality.
			In company with \gls{bcd}, we also plotted the convergence results of \gls{kkt} and \gls{pga} algorithms in Fig.~\ref{fg:wsr_convergence} to show how much performance is gained by solving each subproblem.
			It is observed that Algorithm \ref{al:input_distribution} and \ref{al:active_beamforming} take around \num{100} and \num{10} iterations to converge, respectively.
			Overall, the \gls{bcd} algorithm requires at most \num{5} iterations to converge.
			As $\rho$ increases (not presented here), the convergence of all three algorithms are much faster thanks to the progressive initialization.
		\end{subsubsection}
	\end{subsection}

	\begin{subsection}{Comparison of Scattering Applications}
		On top of the setup in Fig.~\ref{fg:riscatter_network}, we consider RIScatter and the following benchmark applications:
		\begin{itemize}
			\item \emph{Legacy:} Legacy transmission without scatter nodes.
			\item \emph{\gls{bbc}:} The primary signal is \gls{cw} and the receive signal is
			\begin{equation}
				y^{\text{\gls{bbc}}}[n] = \Bigl(\boldsymbol{h}_{\text{D}}^\mathsf{H} + \sum_{k} \alpha_k \boldsymbol{h}_{\text{C},k}^\mathsf{H} x_k\Bigr) \boldsymbol{w} + v[n].
			\end{equation}
			The total backscatter rate approaches $K \log M$ when $N$ is sufficiently large.
			\item \emph{\gls{ambc}:} The user decodes each link independently and semi-coherently while treating the other as interference.
			The primary achievable rate is approximately\footnote{The scattered component is treated as interference with average power $\mathbb{E}\bigl\{\sum_{k} \alpha_k \boldsymbol{h}_{\text{C},k}^\mathsf{H} x_k \boldsymbol{w}s[n]\bigr\} = \sum_{k} \lvert \alpha_k \boldsymbol{h}_{\text{C},k}^\mathsf{H} \boldsymbol{w} \rvert^2$ \cite{Long2020a}.\label{fn:ambc}}
			\begin{equation}
				I_{\text{P}}^{\text{\gls{ambc}}}(s;y) \approx \log \Bigl(1 + \frac{\lvert\boldsymbol{h}_{\text{D}}^\mathsf{H}\boldsymbol{w}\rvert^2}{\sum_{k}\lvert \alpha_k \boldsymbol{h}_{\text{C},k}^\mathsf{H} \boldsymbol{w}\rvert^2+\sigma_v^2}\Bigr),
			\end{equation}
			while the total backscatter rate follows \eqref{eq:backscatter_mutual_information} with uniform input distribution.
			\item \emph{\gls{sr}:} For a sufficiently large $N$, the average primary rate under semi-coherent detection asymptotically approaches \eqref{eq:primary_mutual_information} with uniform input distribution \cite{Long2020a}.
			When $s[n]$ is successfully decoded and the direct interference $\boldsymbol{h}_{\text{D}}^\mathsf{H} \boldsymbol{w} s[n]$ is perfectly cancelled, the intermediate signal is
			\begin{equation}
				\hat{y}^{\text{\gls{sr}}}[n] = \sum_{k} \alpha_k \boldsymbol{h}_{\text{C},k}^\mathsf{H} x_k \boldsymbol{w} s[n] + v[n].
				\label{eq:intermediate_signal}
			\end{equation}
			The total backscatter rate approaches $K \log M$.
			\item \emph{\gls{ris}:} The reflection pattern is deterministic and the total backscatter rate is zero.
			The primary achievable rate is a special case of \eqref{eq:primary_mutual_information}
			\begin{equation}
				I_{\text{P}}^{\text{\gls{ris}}}(s;y|x_{\mathcal{K}}) = I_{\text{P}}(s;y|x_{m_{\mathcal{K}}^{\star}}) = \log \Bigl(1 + \frac{\lvert \boldsymbol{h}^\mathsf{H}(x_{m_{\mathcal{K}}^{\star}}) \boldsymbol{w} \rvert^2}{\sigma_v^2}\Bigr),
			\end{equation}
			where $m_{\mathcal{K}}^{\star} = \arg \max_{m_{\mathcal{K}}} I_{\text{P}}^{\text{\gls{ris}}}(s;y|x_{m_{\mathcal{K}}})$.
		\end{itemize}
		\begin{figure}[!t]
			\centering
			\resizebox{0.65\columnwidth}{!}{
%
%
\definecolor{mycolor1}{rgb}{0.30100,0.74500,0.93300}%
\definecolor{mycolor2}{rgb}{0.46600,0.67400,0.18800}%
\definecolor{mycolor3}{rgb}{0.49400,0.18400,0.55600}%
\definecolor{mycolor4}{rgb}{0.92900,0.69400,0.12500}%
\definecolor{mycolor5}{rgb}{0.85000,0.32500,0.09800}%
\definecolor{mycolor6}{rgb}{0.00000,0.44700,0.74100}%
\begin{tikzpicture}

\begin{axis}[%
width=4.014in,
height=1.552in,
at={(0.673in,0.356in)},
scale only axis,
xmin=0,
xmax=6.5227548374066,
xlabel style={font=\color{white!15!black}},
xlabel={Primary Rate [bits/s/Hz]},
ymin=0,
ymax=2,
ylabel style={font=\color{white!15!black}},
ylabel={Backscatter\\Rate [bits/BB]},
axis background/.style={fill=white},
xmajorgrids,
ymajorgrids,
legend style={at={(0.03,0.97)}, anchor=north west, legend cell align=left, align=left, draw=white!15!black},
align=center,
title style={font=\LARGE},
label style={font=\LARGE},
ticklabel style={font=\Large},
legend style={font=\Large},
reverse legend,
every axis plot/.append style={line width=2pt}
]
\addplot [color=mycolor1, line width=2.0pt, mark=triangle, mark options={solid, rotate=180, mycolor1}]
  table[row sep=crcr]{%
6.33724402501803	1.39764296268229\\
0	1.39764296268229\\
0	0\\
6.52275483678685	0\\
6.52275483678685	2.37312052924553e-08\\
6.38739454470992	1.3235811846854\\
6.35422040211439	1.38916612409166\\
6.34857862205488	1.39386686170527\\
6.3445353675389	1.39608085009443\\
6.34291602910554	1.3966977322109\\
6.34149846039129	1.39711118534274\\
6.34024731759122	1.39737797077851\\
6.3391350240222	1.3975379071887\\
6.33862394227692	1.39758701988787\\
6.33813974610465	1.39761939115801\\
6.33795313341839	1.39762818964178\\
6.33777036860279	1.39763482338317\\
6.3375913342378	1.3976394187338\\
6.33741590881855	1.39764209463684\\
6.33724402501803	1.39764296268229\\
};
\addlegendentry{RIScatter}

\addplot[only marks, mark=triangle, mark options={}, mark size=2.3570pt, draw=mycolor2] table[row sep=crcr]{%
x	y\\
6.5227548374066	0\\
};
\addlegendentry{RIS}

\addplot[only marks, mark=+, mark options={}, mark size=3.5355pt, draw=mycolor3] table[row sep=crcr]{%
x	y\\
6.34321982467796	2\\
};
\addlegendentry{SR}

\addplot[only marks, mark=x, mark options={}, mark size=3.5355pt, draw=mycolor4] table[row sep=crcr]{%
x	y\\
5.90778796357092	1.34857377961699\\
};
\addlegendentry{AmBC}

\addplot[only marks, mark=square, mark options={}, mark size=2.5000pt, draw=mycolor5] table[row sep=crcr]{%
x	y\\
0	1.99999999999997\\
};
\addlegendentry{BBC}

\addplot[only marks, mark=o, mark options={}, mark size=2.7386pt, draw=mycolor6] table[row sep=crcr]{%
x	y\\
6.34314881160129	0\\
};
\addlegendentry{Legacy}

\end{axis}
\end{tikzpicture}%
			}
			\caption{Typical achievable rate region/points of scattering applications for $Q=1$, $K=1$, $M=4$, $N=\num{e3}$, $\sigma_v^2=\qty{-40}{dBm}$ and $r=\qty{2}{m}$.}
			\label{fg:region_comparison}
		\end{figure}
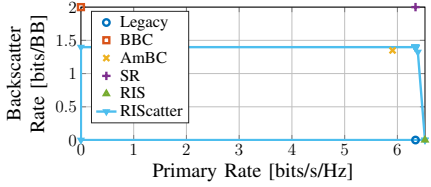

		Fig.~\ref{fg:region_comparison} compares the typical achievable rate region/points of RIScatter and those strategies.
		\emph{First,} we observe \gls{bbc} and \gls{sr} achieve the best backscatter performance thanks to the coherent decoding.
		For \gls{sr}, this comes with the cost of $N$ re-encoding, precoding, subtraction together with a time-domain \gls{mrc} per backscatter symbol.
		Since \gls{sr} requires a very large $N$ to guarantee the primary rate, the signal processing cost at the receiver can be prohibitive, and the backscatter \emph{throughput} can be severely constrained.
		\emph{Second,} the average primary rate slightly decreases/increases in the presence of a \gls{ambc}/\gls{ris} node, and the benefit of \gls{sr} is not obvious.
		This is because the cascaded channel is around \qty{25}{dB} weaker than the direct channel.
		Here, \gls{ris} ensures a constructive superposition of the direct and scattered components, while \gls{sr} only creates a quasi-static rich-scattering environment that marginally enhances the average primary rate.
		When $N$ is moderate, the randomly scattered signals should be modelled as interference rather than stable multipath, and the \gls{sr} point should move vertically towards the \gls{ambc} point.
		\emph{Third,} RIScatter enables a flexible primary-backscatter tradeoff with adaptive input distribution design.
		In terms of maximum primary achievable rate, RIScatter coincides with \gls{ris} and outperforms the others by using the static reflection pattern that maximizes the primary \gls{snr} all the time.
		On the other hand, its maximum backscatter achievable rate is higher than that of \gls{ambc}.
		This is because the adaptive channel coding of RIScatter outperforms the equiprobable inputs of \gls{ambc}, especially at the low backscatter \gls{snr} caused by double fading.
		When multiple antenna is available at the \gls{ap}, active beamforming can be optimized for RIScatter nodes and the advantage over \gls{ambc} should be more prominent.
	\end{subsection}

	\begin{subsection}{Input Distribution under Different \gls{qos}}
		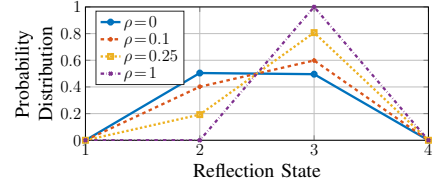
\begin{figure}[!t]
			\centering
			\resizebox{0.65\columnwidth}{!}{
%
%
\definecolor{mycolor1}{rgb}{0.00000,0.44706,0.74118}%
\definecolor{mycolor2}{rgb}{0.85098,0.32549,0.09804}%
\definecolor{mycolor3}{rgb}{0.92941,0.69412,0.12549}%
\definecolor{mycolor4}{rgb}{0.49412,0.18431,0.55686}%
\begin{tikzpicture}

\begin{axis}[%
width=4.079in,
height=1.587in,
at={(0.684in,0.361in)},
scale only axis,
xmin=1,
xmax=4,
xtick={1, 2, 3, 4},
xlabel style={font=\color{white!15!black}},
xlabel={Reflection State},
ymin=0,
ymax=1,
ytick={  0, 0.2, 0.4, 0.6, 0.8,   1},
ylabel style={font=\color{white!15!black}},
ylabel={Probability\\Distribution},
axis background/.style={fill=white},
xmajorgrids,
ymajorgrids,
legend style={at={(0.03,0.97)}, anchor=north west, legend cell align=left, align=left, draw=white!15!black},
align=center,
title style={font=\LARGE},
label style={font=\LARGE},
ticklabel style={font=\Large},
legend style={font=\Large}
]
\addplot [color=mycolor1, line width=2.0pt, mark=o, mark options={solid, mycolor1}]
  table[row sep=crcr]{%
1	6.48337921881147e-07\\
2	0.504332536970115\\
3	0.495666065268481\\
4	7.49423481629768e-07\\
};
\addlegendentry{$\rho =0$}

\addplot [color=mycolor2, dashed, line width=2.0pt, mark=+, mark options={solid, mycolor2}]
  table[row sep=crcr]{%
1	9.07520883479571e-08\\
2	0.401497967775065\\
3	0.598501822229458\\
4	1.1924338950349e-07\\
};
\addlegendentry{$\rho =0.1$}

\addplot [color=mycolor3, dotted, line width=2.0pt, mark=square, mark options={solid, mycolor3}]
  table[row sep=crcr]{%
1	5.20790909800095e-09\\
2	0.192484685585801\\
3	0.807515300031504\\
4	9.17478600370393e-09\\
};
\addlegendentry{$\rho =0.25$}

\addplot [color=mycolor4, dashdotted, line width=2.0pt, mark=x, mark options={solid, mycolor4}]
  table[row sep=crcr]{%
1	1.08446044988811e-11\\
2	8.27854203489531e-06\\
3	0.999991721175464\\
4	2.71656646770891e-10\\
};
\addlegendentry{$\rho =1$}

\end{axis}
\end{tikzpicture}%
			}
			\caption{Typical RIScatter reflection state distribution at different $\rho$ for $Q=1$, $K=1$, $M=4$, $N=20$, $\sigma_v^2=\qty{-40}{dBm}$ and $r=\qty{2}{m}$.}
			\label{fg:distribution_weights}
		\end{figure}
		The objective is to demonstrate RIScatter nodes can leverage \gls{csi}- and \gls{qos}-adaptive input distribution design to balance backscatter modulation and passive beamforming.
		For one RIScatter node with $M=4$, we evaluate the \gls{kkt} input distribution at different \gls{qos} and present the result in Fig.~\ref{fg:distribution_weights}.
		At $\rho=0$ where the backscatter performance is prioritized, the optimal input distribution is \num{0} on two states and nearly uniform on the other two.
		This is inline with Shannon's observation that binary antipodal inputs is good enough for channel capacity at low \gls{snr} \cite{Shannon1948}.
		When the scattered signal is relatively weak, the conditional energy \gls{pdf} under different hypotheses can be closely spaced as in Fig.~\ref{fg:energy_distribution}.
		The extreme states producing the lowest/highest energy are always assigned with non-zero probability, while the middles cannot provide enough energy diversity and end up unused.
		At $\rho=1$ where the primary link is prioritized, the optimal input distribution is $[0, 0, 1, 0]^\mathsf{T}$ since state 3 provides higher primary \gls{snr} than other states.
		That is, the reflection pattern becomes deterministic and the RIScatter node boils down to a static discrete \gls{ris} element.
		Increasing $\rho$ from \num{0} to \num{1} creates a smooth transition from backscatter modulation to passive beamforming, suggesting RIScatter unifies \gls{bc} and \gls{ris} from a probabilistic perspective.
	\end{subsection}

	\begin{subsection}{Rate Region by Different Schemes}
		\begin{figure}[!t]
			\centering
			\subfloat[Input Distribution, $Q=1$\label{fg:region_distribution}]{
				\resizebox{0.6\columnwidth}{!}{
					\input{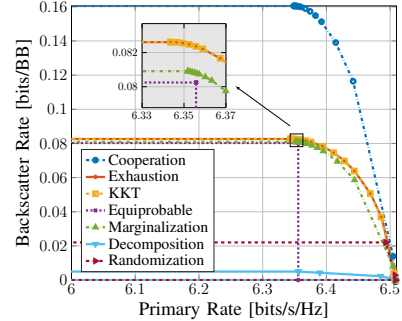}
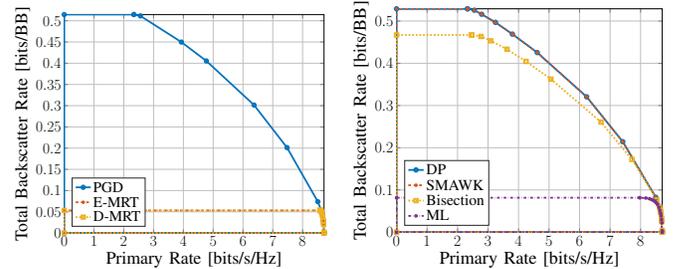
				}
			}
			\\
			\subfloat[Active Beamforming, $Q=4$\label{fg:region_beamforming}]{
				\resizebox{0.48\columnwidth}{!}{
%
%
\definecolor{mycolor1}{rgb}{0.00000,0.44706,0.74118}%
\definecolor{mycolor2}{rgb}{0.85098,0.32549,0.09804}%
\definecolor{mycolor3}{rgb}{0.92941,0.69412,0.12549}%
\begin{tikzpicture}

\begin{axis}[%
width=4.079in,
height=3.432in,
at={(0.684in,0.463in)},
scale only axis,
xmin=0,
xmax=8.71998560599984,
xlabel style={font=\color{white!15!black}},
xlabel={Primary Rate [bits/s/Hz]},
ymin=0,
ymax=0.514663084739105,
ylabel style={font=\color{white!15!black}},
ylabel={Total Backscatter Rate [bits/BB]},
axis background/.style={fill=white},
xmajorgrids,
ymajorgrids,
legend style={at={(0.03,0.03)}, anchor=south west, legend cell align=left, align=left, draw=white!15!black},
title style={font=\huge},
label style={font=\huge},
ticklabel style={font=\LARGE},
legend style={font=\LARGE},
scaled y ticks=false,
y tick label style={/pgf/number format/.cd, fixed, precision=2}
]
\addplot [color=mycolor1, line width=2.0pt, mark=o, mark options={solid, mycolor1}]
  table[row sep=crcr]{%
2.33916065404713	0.514663084739105\\
0	0.514663084739105\\
0	0\\
8.71877007451367	0\\
8.71877007451367	7.74873204371495e-08\\
8.71805205557953	0.00244067485556787\\
8.70591606075502	0.02104932180974\\
8.69511588952171	0.0297435430193558\\
8.67624130058129	0.0396481462824668\\
8.66228083359227	0.0449052199961776\\
8.6444096988787	0.0499695917150902\\
8.51256151944983	0.073749445384341\\
7.47793088269382	0.201239566488916\\
6.37096716809068	0.301510590669739\\
4.77132245802858	0.405335141657174\\
3.93359101103267	0.449788833316287\\
2.55872317562845	0.511432482879822\\
2.33916065404713	0.514663084739105\\
};
\addlegendentry{PGD}

\addplot [color=mycolor2, dashed, line width=2.0pt, mark=+, mark options={solid, mycolor2}]
  table[row sep=crcr]{%
8.59178340066455	0.0532597352790397\\
0	0.0532597352790397\\
0	0\\
8.71998560599984	0\\
8.71998560599984	7.73794332514691e-08\\
8.71929985295599	0.00243856589119924\\
8.70745638508896	0.0206762162023952\\
8.69728566162353	0.0287174278389967\\
8.67998815304844	0.037641785235532\\
8.66735227066505	0.0422724349391682\\
8.652677785857	0.046379382766988\\
8.63656254444798	0.0496443071521638\\
8.61975942656468	0.0518786562946866\\
8.61152265624827	0.052577725571065\\
8.60357392125354	0.0530212092775739\\
8.60049442361505	0.0531315600842072\\
8.59750390999725	0.0532068135371646\\
8.59457596038297	0.0532488928934199\\
8.59178340066455	0.0532597352790397\\
};
\addlegendentry{E-MRT}

\addplot [color=mycolor3, dotted, line width=2.0pt, mark=square, mark options={solid, mycolor3}]
  table[row sep=crcr]{%
8.59153976062595	0.0533210645407662\\
0	0.0533210645407662\\
0	0\\
8.71871799332824	0\\
8.71871799332824	7.78893492328268e-08\\
8.71808135290584	0.00225351571682053\\
8.70661729288011	0.0200596524211706\\
8.69655053848953	0.0281803606664607\\
8.67958362885197	0.0371191448678512\\
8.66727216045506	0.0417190220993342\\
8.65273435931506	0.0458890069507623\\
8.63664297257043	0.049269565030704\\
8.61977377622263	0.0516499368085544\\
8.61150467986663	0.0524300531488041\\
8.60344696735477	0.0529562915270052\\
8.60034982199742	0.0530991303439398\\
8.59733768667359	0.0532061823236453\\
8.59436568598963	0.0532792867423072\\
8.59153976062595	0.0533210645407662\\
};
\addlegendentry{D-MRT}

\end{axis}
\end{tikzpicture}%
				}
			}
			\subfloat[Decision Threshold, $Q=4$\label{fg:region_threshold}]{
				\resizebox{0.48\columnwidth}{!}{
%
%
\definecolor{mycolor1}{rgb}{0.00000,0.44706,0.74118}%
\definecolor{mycolor2}{rgb}{0.85098,0.32549,0.09804}%
\definecolor{mycolor3}{rgb}{0.92941,0.69412,0.12549}%
\definecolor{mycolor4}{rgb}{0.49412,0.18431,0.55686}%
\begin{tikzpicture}

\begin{axis}[%
width=4.079in,
height=3.432in,
at={(0.684in,0.463in)},
scale only axis,
xmin=0,
xmax=8.71037769136309,
xlabel style={font=\color{white!15!black}},
xlabel={Primary Rate [bits/s/Hz]},
ymin=0,
ymax=0.529009349687418,
ylabel style={font=\color{white!15!black}},
ylabel={Total Backscatter Rate [bits/BB]},
axis background/.style={fill=white},
xmajorgrids,
ymajorgrids,
legend style={at={(0.03,0.03)}, anchor=south west, legend cell align=left, align=left, draw=white!15!black},
title style={font=\huge},
label style={font=\huge},
ticklabel style={font=\LARGE},
legend style={font=\LARGE},
scaled y ticks=false,
y tick label style={/pgf/number format/.cd, fixed, precision=2}
]
\addplot [color=mycolor1, line width=2.0pt, mark=o, mark options={solid, mycolor1}]
  table[row sep=crcr]{%
2.32665775109935	0.529009349687418\\
0	0.529009349687418\\
0	0\\
8.70937198183187	0\\
8.70937198183187	7.69303579579504e-08\\
8.70812675507196	0.00424644852742044\\
8.69434563757341	0.025573983576851\\
8.68282764822907	0.0348822939505003\\
8.66211786962418	0.0458251586048386\\
8.64487268552619	0.0524344721719585\\
8.6147169918115	0.0612134141401668\\
8.4978742152146	0.0826575883446601\\
7.41652376108267	0.214293851916817\\
6.23129272558466	0.320668859458555\\
4.61172545043024	0.42545971377814\\
3.79811459993038	0.469040566793437\\
3.2422041362473	0.496918273956616\\
2.79071942475067	0.516115642119426\\
2.55141194115854	0.525628345755559\\
2.32665775109935	0.529009349687418\\
};
\addlegendentry{DP}

\addplot [color=mycolor2, dashed, line width=2.0pt, mark=+, mark options={solid, mycolor2}]
  table[row sep=crcr]{%
2.32665775109935	0.529009349687418\\
0	0.529009349687418\\
0	0\\
8.70937198183187	0\\
8.70937198183187	7.69303579579504e-08\\
8.70812675507196	0.00424644852742044\\
8.69434563757341	0.025573983576851\\
8.68282764822907	0.0348822939505003\\
8.66211786962418	0.0458251586048386\\
8.64487268552619	0.0524344721719585\\
8.6147169918115	0.0612134141401668\\
8.4978742152146	0.0826575883446601\\
7.41652376108267	0.214293851916817\\
6.23129272558466	0.320668859458555\\
4.61172545043024	0.42545971377814\\
3.79811459993038	0.469040566793437\\
3.2422041362473	0.496918273956616\\
2.79071942475067	0.516115642119426\\
2.55141194115854	0.525628345755559\\
2.32665775109935	0.529009349687418\\
};
\addlegendentry{SMAWK}

\addplot [color=mycolor3, dotted, line width=2.0pt, mark=square, mark options={solid, mycolor3}]
  table[row sep=crcr]{%
2.46146128286634	0.466990714058083\\
0	0.466990714058083\\
0	0\\
8.7095354833513	0\\
8.7095354833513	7.42357744836325e-08\\
8.70843124696874	0.00397027637710423\\
8.69492401613322	0.0249877099860776\\
8.68308680890842	0.0345194229476592\\
8.66256035576318	0.0455398120759765\\
8.6450287539251	0.0522717497640591\\
8.61915873481432	0.0598903492121913\\
8.53212717932426	0.0761292782838948\\
7.71671525715228	0.172225370611815\\
6.70222882861717	0.260854612088341\\
5.06021700037469	0.362503137063897\\
4.24652763561364	0.404443983812365\\
3.62017725533614	0.433280503381125\\
3.08584299799995	0.453226982210376\\
2.77091480619372	0.463468311567622\\
2.46146128286634	0.466990714058083\\
};
\addlegendentry{Bisection}

\addplot [color=mycolor4, dashdotted, line width=2.0pt, mark=x, mark options={solid, mycolor4}]
  table[row sep=crcr]{%
7.97257156626925	0.0815980199067495\\
0	0.0815980199067495\\
0	0\\
8.71037769136309	0\\
8.71037769136309	5.93897901276753e-08\\
8.70941126476681	0.00344299828529007\\
8.69761552618312	0.0219596832111129\\
8.68786841894045	0.0298809304919299\\
8.67092053432189	0.0391127961080852\\
8.65563531378827	0.0450685642534716\\
8.63223262018689	0.0520130694335445\\
8.60011739459726	0.0590984559662197\\
8.55257482814744	0.0659146527946499\\
8.5132847820689	0.069772824652909\\
8.46536377449469	0.0731121533451646\\
8.42465045898258	0.0749718059283308\\
8.36247374429884	0.0772982944891839\\
8.29519095697419	0.0790947770434924\\
8.17672391400642	0.080822334320136\\
7.97257156626925	0.0815980199067495\\
};
\addlegendentry{ML}

\end{axis}
\end{tikzpicture}%
				}
			}
			\caption{
				Average primary-total-backscatter rate regions by different input distribution, active beamforming, and decision threshold schemes for $K=2$, $M=4$, $N=20$, $\sigma_v^2=\qty{-40}{dBm}$ and $r=\qty{2}{m}$.
			}
		\end{figure}
		\begin{subsubsection}{Input Distribution}
			We compare these input distribution designs for problem \eqref{op:input_distribution}:
			\begin{itemize}
				\item \emph{Cooperation:} Joint encoding using a joint probability array $p(x_{m_{\mathcal{K}}})$ with $M^K$ entries by Algorithm \ref{al:input_distribution};
				\item \emph{Exhaustion:} Exhaustive search over the $M$-dimensional probability simplex with resolution $\Delta p = \num{e-2}$;
				\item \emph{\gls{kkt}:} Solution by Algorithm \ref{al:input_distribution};
				\item \emph{Equiprobable:} Uniform input distribution.
			\end{itemize}
			We also consider these independent distribution recovery methods from the joint probability array:
			\begin{itemize}
				\item \emph{Marginalization:} Marginal probability distributions;
				\item \emph{Decomposition:} Normalized rank-\num{1} \gls{cp} decomposed tensors by \texttt{Tensor Toolbox} \cite{Bader2022};
				\item \emph{Randomization:} Gaussian randomization with the guidance of correlation matrix \cite{Calvo2010}.
			\end{itemize}

			Fig.~\subref*{fg:region_distribution} shows their average achievable rate regions.
			Cooperation achieves the outer bound of all schemes, but joint encoding over passive devices may incur additional hardware cost.
			Besides, the average rate performance of Exhaustion and \gls{kkt} completely coincide with each other when $K=2$.
			This confirms Remark \ref{re:input_kkt_distribution} that the \gls{kkt} input distribution can be good enough when $K$ is moderate.
			Equiprobable experiences minor backscatter and major primary rate losses without exploiting \gls{csi} and \gls{qos}, and those gaps should be larger when $M$ or $K$ increases.
			Marginalization provides a close performance to \gls{kkt}, but Randomization and Decomposition fail our expectations for most channel realizations.
			Those observations emphasize the importance of adaptive RIScatter encoding and demonstrate the advantage of the proposed \gls{kkt} input distribution design.
		\end{subsubsection}

		\begin{subsubsection}{Active Beamforming}
			We consider three typical active beamforming schemes for problem \eqref{op:active_beamforming}:
			\begin{itemize}
				\item \emph{\gls{pga}:} Solution by Algorithm \ref{al:active_beamforming};
				\item \emph{E-\gls{mrt}:} \gls{mrt} towards the ergodic composite channel $\sum_{m_{\mathcal{K}}} p(x_{m_{\mathcal{K}}}) \boldsymbol{h}^\mathsf{H}(x_{m_{\mathcal{K}}})$;
				\item \emph{D-\gls{mrt}:} \gls{mrt} towards the direct channel $\boldsymbol{h}_{\text{D}}^\mathsf{H}$.
			\end{itemize}

			Fig.~\subref*{fg:region_beamforming} presents the average achievable rate regions for those schemes.
			In the low-$\rho$ regime, the proposed \gls{pga} beamformer significantly outperforms both \gls{mrt} schemes in terms of total backscatter rate.
			This is because the semi-coherent backscatter decoding relies on the relative energy difference under different backscatter symbol tuples.
			Such an energy diversity is enhanced by \gls{pga} that effectively exploits backscatter constellation and input distribution knowledge rather than simply maximizes the channel strength.
			As $\rho$ increases, the primary \gls{snr} outweighs the backscatter energy difference in \eqref{eq:weighted_sum_mutual_information_explicit}, and \gls{pga} beamformer approaches E-\gls{mrt}.
			At $\rho=1$, both \gls{pga} and E-\gls{mrt} boil down to \gls{mrt} towards the strongest composite channel.
			The difference between E-\gls{mrt} and D-\gls{mrt} is insignificant when RIScatter nodes are dispersed.
			Those observations confirm that the proposed \gls{pga} active beamforming design can exploit the \gls{csi}, \gls{qos}, and backscatter constellation to enlarge the achievable rate region.
		\end{subsubsection}

		\begin{subsubsection}{Decision Threshold}
			We evaluate the following decision threshold strategies for problem \eqref{op:decision_threshold_discrete}:
			\begin{itemize}
				\item \emph{\gls{dp}:} Benchmark \gls{dp} method for sequential quantizer \cite{He2021};
				\item \emph{\gls{smawk}:} \emph{\gls{dp}} accelerated by the \gls{smawk} algorithm \cite{He2021};
				\item \emph{Bisection:} The traverse-then-bisect algorithm \cite{Nguyen2020a};
				\item \emph{\gls{ml}:} Maximum likelihood detector \eqref{eq:detection_threshold_ml} \cite{Qian2019}.
			\end{itemize}

			Fig.~\subref*{fg:region_threshold} reveals the average achievable rate region for those strategies.
			The distribution-aware schemes \gls{dp}, \gls{smawk} and Bisection ensure higher total backscatter rate than \gls{ml}.
			This is because the total backscatter rate \eqref{eq:backscatter_mutual_information} is a function of both input distribution and decision regions, and the rate-optimal thresholding depends heavily on the input distribution.
			For example, the backscatter symbol tuples with zero input probability should be assigned with empty decision regions, in order to increase the success detection rates of other hypotheses.
			It highlights the importance of joint input distribution and decision threshold design in rate maximization problems.
		\end{subsubsection}

	\end{subsection}

	\begin{subsection}{Rate Region under Different Configurations}
		\begin{figure}[!t]
			\centering
			\subfloat[RIScatter Nodes\label{fg:region_tags}]{
				\resizebox{0.48\columnwidth}{!}{
%
%
\definecolor{mycolor1}{rgb}{0.00000,0.44706,0.74118}%
\definecolor{mycolor2}{rgb}{0.85098,0.32549,0.09804}%
\definecolor{mycolor3}{rgb}{0.92941,0.69412,0.12549}%
\definecolor{mycolor4}{rgb}{0.49412,0.18431,0.55686}%
\begin{tikzpicture}

\begin{axis}[%
width=4.014in,
height=3.397in,
at={(0.673in,0.459in)},
scale only axis,
xmin=0,
xmax=8.84663715678527,
xlabel style={font=\color{white!15!black}},
xlabel={Primary Rate [bits/s/Hz]},
ymin=0,
ymax=0.80209549735363,
ylabel style={font=\color{white!15!black}},
ylabel={Total Backscatter Rate [bits/BB]},
axis background/.style={fill=white},
xmajorgrids,
ymajorgrids,
legend style={legend cell align=left, align=left, draw=white!15!black},
title style={font=\huge},
label style={font=\huge},
ticklabel style={font=\LARGE},
legend style={font=\LARGE},
scaled y ticks=false,
y tick label style={/pgf/number format/.cd, fixed, precision=2}
]
\addplot [color=mycolor1, line width=2.0pt, mark=o, mark options={solid, mycolor1}]
  table[row sep=crcr]{%
8.22667029994347	0.017540563142412\\
0	0.017540563142412\\
0	0\\
8.64369877031669	0\\
8.64369877031669	5.22435704533235e-08\\
8.64361724690434	0.000273700486979121\\
8.64132200711325	0.00377019587032916\\
8.6393656533708	0.00537761652756264\\
8.63587952709457	0.00727793290173079\\
8.6330279141891	0.00839415414221361\\
8.62865709669078	0.00969862961396924\\
8.62142817711826	0.0112773570774465\\
8.61163461307148	0.0127441574048236\\
8.60702261357685	0.0132407114579893\\
8.59313101408756	0.01417811001091\\
8.58554286372929	0.0145768863951287\\
8.57535331632231	0.0149948903875726\\
8.55478197113172	0.0155658634490472\\
8.49947538245987	0.0164376399821914\\
8.22667029994347	0.017540563142412\\
};
\addlegendentry{$K =1$}

\addplot [color=mycolor2, dashed, line width=2.0pt, mark=+, mark options={solid, mycolor2}]
  table[row sep=crcr]{%
8.0745496198252	0.0389423160644722\\
0	0.0389423160644722\\
0	0\\
8.67414921026067	0\\
8.67414921026067	7.11479478884334e-08\\
8.67389584123813	0.000869934851270209\\
8.66830360735917	0.00930437026622899\\
8.66359875313575	0.0131183738983091\\
8.6543529696745	0.017863506108297\\
8.64817287903989	0.0202688856897266\\
8.6394578652119	0.0228993209047608\\
8.62498321643782	0.0260565615218228\\
8.60039463621136	0.0295646137113563\\
8.58636756921059	0.0310377997463991\\
8.5664845527562	0.0325611282770937\\
8.55384762687453	0.0332673264669256\\
8.51996834792101	0.0345728071987884\\
8.47453422086977	0.0358480703769788\\
8.35574115651914	0.0376902259259884\\
8.0745496198252	0.0389423160644722\\
};
\addlegendentry{$K =2$}

\addplot [color=mycolor3, dotted, line width=2.0pt, mark=square, mark options={solid, mycolor3}]
  table[row sep=crcr]{%
2.31987036642052	0.517539967309673\\
0	0.517539967309673\\
0	0\\
8.73322374429886	0\\
8.73322374429886	7.73874792352331e-08\\
8.73228869013851	0.00331530245765858\\
8.71922310288744	0.0235917628903182\\
8.7088648257655	0.0319895639740332\\
8.68796080105943	0.042757378309304\\
8.67351187415625	0.0481378217796802\\
8.65250968695428	0.0542547589049583\\
8.5361579638998	0.0751466470119891\\
7.45459781206878	0.206063931086682\\
6.35918383876333	0.304232634411087\\
4.722320926516	0.411025776691365\\
3.98533522602161	0.451330121965152\\
2.80296564868843	0.504550240560943\\
2.55158717999824	0.514124185285439\\
2.31987036642052	0.517539967309673\\
};
\addlegendentry{$K =4$}

\addplot [color=mycolor4, dashdotted, line width=2.0pt, mark=x, mark options={solid, mycolor4}]
  table[row sep=crcr]{%
1.63718483875947	0.80209549735363\\
0	0.80209549735363\\
0	0\\
8.84663715678527	0\\
8.84663715678527	7.54410878790313e-08\\
8.84514791257237	0.00524052776761055\\
8.82164841110655	0.0417019163784306\\
8.80096091577627	0.0581183463830854\\
8.76337637533374	0.0775856402558251\\
8.73263250077828	0.0887860749141904\\
8.67834842677259	0.103731120164347\\
8.06417324399172	0.210286899800982\\
5.8856705018099	0.483533819727886\\
4.19928212203302	0.640269773281524\\
2.43119266684432	0.777441530723861\\
2.21932852000179	0.792798740305588\\
2.10810114956683	0.799294292741716\\
1.63718483875947	0.80209549735363\\
};
\addlegendentry{$K =8$}

\end{axis}
\end{tikzpicture}%
				}
			}
			\subfloat[Reflection States\label{fg:region_states}]{
				\resizebox{0.48\columnwidth}{!}{
%
%
\definecolor{mycolor1}{rgb}{0.00000,0.44706,0.74118}%
\definecolor{mycolor2}{rgb}{0.85098,0.32549,0.09804}%
\definecolor{mycolor3}{rgb}{0.92941,0.69412,0.12549}%
\definecolor{mycolor4}{rgb}{0.49412,0.18431,0.55686}%
\begin{tikzpicture}

\begin{axis}[%
width=4.079in,
height=3.432in,
at={(0.684in,0.463in)},
scale only axis,
xmin=0,
xmax=8.65005050060273,
xlabel style={font=\color{white!15!black}},
xlabel={Primary Rate [bits/s/Hz]},
ymin=0,
ymax=0.159990686232269,
ylabel style={font=\color{white!15!black}},
ylabel={Backscatter Rate [bits/BB]},
axis background/.style={fill=white},
xmajorgrids,
ymajorgrids,
legend style={legend cell align=left, align=left, draw=white!15!black},
title style={font=\huge},
label style={font=\huge},
ticklabel style={font=\LARGE},
legend style={font=\LARGE},
scaled y ticks=false,
y tick label style={/pgf/number format/.cd, fixed, precision=2}
]
\addplot [color=mycolor1, line width=2.0pt, mark=o, mark options={solid, mycolor1}]
  table[row sep=crcr]{%
8.21313583828618	0.0161303040729969\\
0	0.0161303040729969\\
0	0\\
8.6380164413452	0\\
8.6380164413452	5.23335067600407e-08\\
8.63798532661636	0.000103463079183655\\
8.63619983785027	0.00266485830082033\\
8.63435267606119	0.00419126279589298\\
8.63118735148684	0.00590094582937064\\
8.6284371575562	0.00696695466588222\\
8.6241962841021	0.00823317473169911\\
8.61736295042584	0.00971846873210729\\
8.60706281141783	0.0112618749480944\\
8.59985036384004	0.0119852467071752\\
8.59006180884157	0.0127067337316147\\
8.58434938574194	0.0130226471035682\\
8.57180983896197	0.0135190262801698\\
8.55280973893037	0.0140577968301156\\
8.49642457782116	0.0149475892805674\\
8.21313583828618	0.0161303040729969\\
};
\addlegendentry{$M =2$}

\addplot [color=mycolor2, dashed, line width=2.0pt, mark=+, mark options={solid, mycolor2}]
  table[row sep=crcr]{%
8.18642709547471	0.0314121164282678\\
0	0.0314121164282678\\
0	0\\
8.65005049859416	0\\
8.65005049859416	7.09755242417814e-08\\
8.64990915187052	0.000499575972106212\\
8.64531503064754	0.00768957190246738\\
8.64128774734272	0.0110213092071536\\
8.634407183727	0.0148366088220961\\
8.62913329036402	0.0169241046995264\\
8.62094996372438	0.0193877229992253\\
8.60949429743689	0.0219843409167101\\
8.59675032367863	0.0239794730578603\\
8.58890259975571	0.0248537836384125\\
8.57059601179967	0.0262132134946077\\
8.55814588973557	0.0268868297918913\\
8.54071650866786	0.0276045541927192\\
8.50946271259886	0.028522893308445\\
8.41585102209471	0.0300717755553185\\
8.18642709547471	0.0314121164282678\\
};
\addlegendentry{$M =4$}

\addplot [color=mycolor3, dotted, line width=2.0pt, mark=square, mark options={solid, mycolor3}]
  table[row sep=crcr]{%
8.19241172582352	0.0287563704192612\\
0	0.0287563704192612\\
0	0\\
8.64571405988583	0\\
8.64571405988583	7.80949614657515e-08\\
8.64552837670574	0.000663086481332996\\
8.64120711456589	0.00741440717375437\\
8.63755508858577	0.0104314954423075\\
8.63137573848376	0.0138618706054059\\
8.62689616293185	0.015630721147354\\
8.61970375936883	0.0178367687821761\\
8.6102811853774	0.0199959324645335\\
8.599934473816	0.0216794607773852\\
8.59117280659479	0.0226413431937772\\
8.57485653798316	0.0238988077586394\\
8.56577930543271	0.0244179120851519\\
8.54763829543936	0.0251608551295309\\
8.51766225640485	0.0260315190406393\\
8.43347778895146	0.0274568838385641\\
8.19241172582352	0.0287563704192612\\
};
\addlegendentry{$M =8$}

\addplot [color=mycolor4, dashdotted, line width=2.0pt, mark=x, mark options={solid, mycolor4}]
  table[row sep=crcr]{%
2.68444928649785	0.159990686232269\\
0	0.159990686232269\\
0	0\\
8.65005050060273	0\\
8.65005050060273	7.90545196736994e-08\\
8.64983653373986	0.000750849657338434\\
8.64459944873304	0.00908712657315221\\
8.64004961245607	0.0128749387021321\\
8.63246545119624	0.0171202576461528\\
8.6269798064968	0.0192950319253273\\
8.61863651291064	0.0218846962156438\\
8.60550730936448	0.0248844354108822\\
8.58490200055985	0.0280652597079539\\
8.56200510436154	0.0302676265688008\\
8.36872902902114	0.0407934118292221\\
8.00164420579559	0.0555986306411531\\
7.10640116554883	0.084362294171373\\
5.22446207676616	0.130483571966158\\
3.88208318182737	0.151819893311433\\
2.68444928649785	0.159990686232269\\
};
\addlegendentry{$M =16$}

\end{axis}
\end{tikzpicture}%
				}
			}
			\\
			\subfloat[Transmit Antennas\label{fg:region_txs}]{
				\resizebox{0.48\columnwidth}{!}{
%
%
\definecolor{mycolor1}{rgb}{0.00000,0.44706,0.74118}%
\definecolor{mycolor2}{rgb}{0.85098,0.32549,0.09804}%
\definecolor{mycolor3}{rgb}{0.92941,0.69412,0.12549}%
\definecolor{mycolor4}{rgb}{0.49412,0.18431,0.55686}%
\begin{tikzpicture}

\begin{axis}[%
width=4.079in,
height=3.432in,
at={(0.684in,0.463in)},
scale only axis,
xmin=0,
xmax=9.85699759883536,
xlabel style={font=\color{white!15!black}},
xlabel={Primary Rate [bits/s/Hz]},
ymin=0,
ymax=1.04175528860998,
ylabel style={font=\color{white!15!black}},
ylabel={Total Backscatter Rate [bits/BB]},
axis background/.style={fill=white},
xmajorgrids,
ymajorgrids,
legend style={legend cell align=left, align=left, draw=white!15!black},
title style={font=\huge},
label style={font=\huge},
ticklabel style={font=\LARGE},
legend style={font=\LARGE},
scaled y ticks=false,
y tick label style={/pgf/number format/.cd, fixed, precision=2}
]
\addplot [color=mycolor1, line width=2.0pt, mark=o, mark options={solid, mycolor1}]
  table[row sep=crcr]{%
6.42471117417535	0.144376666540991\\
0	0.144376666540991\\
0	0\\
6.74196682868404	0\\
6.74196682868404	5.71741190817808e-08\\
6.74023862813285	0.00610449093268154\\
6.70752703701423	0.0571724435050568\\
6.6783885243017	0.0805352370855534\\
6.63107503428661	0.105474585327752\\
6.59893283473099	0.117444431361871\\
6.56304356786301	0.127719043000317\\
6.52558654234997	0.135584382708212\\
6.48867380480034	0.140829271116383\\
6.47113899510023	0.142496068512134\\
6.45453555457252	0.143594266153564\\
6.44817770528809	0.143890166218654\\
6.44199794342855	0.144111334566488\\
6.43598801452922	0.1442629666094\\
6.43016714717683	0.144349725801049\\
6.42471117417535	0.144376666540991\\
};
\addlegendentry{$Q =1$}

\addplot [color=mycolor2, dashed, line width=2.0pt, mark=+, mark options={solid, mycolor2}]
  table[row sep=crcr]{%
2.79020819303466	0.488195035108493\\
0	0.488195035108493\\
0	0\\
7.75469994971824	0\\
7.75469994971824	6.59519395368363e-08\\
7.75290012945326	0.00639214600401521\\
7.72441958170818	0.0496542237345276\\
7.6997388485625	0.069140650052136\\
7.65701181151499	0.0915154651747935\\
7.62461436589164	0.103730208530772\\
7.57875930230894	0.117110808507555\\
7.39983699105557	0.150861589227721\\
6.44572027958897	0.269406064404286\\
5.53397939237956	0.351192565351045\\
4.54899743357208	0.417053631370612\\
4.12660797540219	0.44304399805258\\
3.76361910460241	0.462869173283531\\
3.31485561242889	0.483534128848625\\
2.79020819303466	0.488195035108493\\
};
\addlegendentry{$Q =2$}

\addplot [color=mycolor3, dotted, line width=2.0pt, mark=square, mark options={solid, mycolor3}]
  table[row sep=crcr]{%
1.69187286204325	0.770512172740724\\
0	0.770512172740724\\
0	0\\
8.84808072388328	0\\
8.84808072388328	7.56839535539499e-08\\
8.84632396812058	0.00630405669348447\\
8.82357251739471	0.0416939365873245\\
8.80337166019166	0.0575947694560284\\
8.76630997139164	0.076726692718367\\
8.73643425545718	0.0876184083460905\\
8.69688515381022	0.098455073787465\\
8.16091941578654	0.19143190342034\\
6.09289744524293	0.451174381463843\\
4.50611881002225	0.598715919102293\\
2.51534899469537	0.745950011580985\\
2.29676590844631	0.761211365163772\\
2.19470188970647	0.767420833964235\\
1.69187286204325	0.770512172740724\\
};
\addlegendentry{$Q =4$}

\addplot [color=mycolor4, dashdotted, line width=2.0pt, mark=x, mark options={solid, mycolor4}]
  table[row sep=crcr]{%
1.48280281555081	1.04175528860998\\
0	1.04175528860998\\
0	0\\
9.85699759883536	0\\
9.85699759883536	8.45330019710809e-08\\
9.85534156209392	0.00598610455161681\\
9.83243544707194	0.0418418490660611\\
9.81418863800443	0.0561725415351849\\
9.77895045160409	0.0737724552557911\\
9.75007397076063	0.0839219072813168\\
9.70765549240287	0.0950077690203414\\
8.59964255380479	0.285804548154435\\
5.16245205048357	0.732612416199533\\
3.47544808618657	0.899228966031378\\
1.98273010368056	1.03639770135034\\
1.902937976093	1.04163127310565\\
1.48280281555081	1.04175528860998\\
};
\addlegendentry{$Q =8$}

\end{axis}
\end{tikzpicture}%
				}
			}
			\subfloat[Spreading Factor\label{fg:region_duration}]{
				\resizebox{0.48\columnwidth}{!}{
%
%
\definecolor{mycolor1}{rgb}{0.00000,0.44706,0.74118}%
\definecolor{mycolor2}{rgb}{0.85098,0.32549,0.09804}%
\definecolor{mycolor3}{rgb}{0.92941,0.69412,0.12549}%
\definecolor{mycolor4}{rgb}{0.49412,0.18431,0.55686}%
\begin{tikzpicture}

\begin{axis}[%
width=4.079in,
height=3.432in,
at={(0.684in,0.463in)},
scale only axis,
xmin=0,
xmax=8.8183272162648,
xlabel style={font=\color{white!15!black}},
xlabel={Primary Rate [bits/s/Hz]},
ymin=0,
ymax=0.0387033207318202,
ylabel style={font=\color{white!15!black}},
ylabel={Total Backscatter Rate [bits/PB]},
axis background/.style={fill=white},
xmajorgrids,
ymajorgrids,
legend style={legend cell align=left, align=left, draw=white!15!black},
title style={font=\huge},
label style={font=\huge},
ticklabel style={font=\LARGE},
legend style={font=\LARGE},
scaled y ticks=false,
y tick label style={/pgf/number format/.cd, fixed, precision=3}
]
\addplot [color=mycolor1, line width=2.0pt, mark=o, mark options={solid, mycolor1}]
  table[row sep=crcr]{%
1.48510320561817	0.0368420449632478\\
0	0.0368420449632478\\
0	0\\
8.81494977830283	0\\
8.81494977830283	3.83377782240587e-09\\
8.81477352076949	4.12086540411783e-05\\
8.80464308481592	0.00150570482545958\\
8.79481987635483	0.00225430896534762\\
8.77713969997226	0.00314972206304367\\
8.76137548769293	0.00370957602960849\\
8.73671299086267	0.00438218933776903\\
8.69453600558026	0.00520601203273033\\
8.28548800580697	0.0100384418588948\\
7.50094640226552	0.0164759402145025\\
6.17365035431449	0.0243306497651073\\
5.38460240649074	0.0281228393432611\\
4.43736577109357	0.0316815244629786\\
3.35788745554615	0.0347239772599961\\
2.42247216478758	0.0363829293680192\\
1.48510320561817	0.0368420449632478\\
};
\addlegendentry{$N =10$}

\addplot [color=mycolor2, dashed, line width=2.0pt, mark=+, mark options={solid, mycolor2}]
  table[row sep=crcr]{%
1.67326087069726	0.0387033207318202\\
0	0.0387033207318202\\
0	0\\
8.81632618502012	0\\
8.81632618502012	3.78776722621619e-09\\
8.81475527355384	0.000280915977176361\\
8.79137070993339	0.00206604004006254\\
8.77128743184392	0.00286700361188652\\
8.73376638330174	0.00384928202287025\\
8.70398959350056	0.00439622330702994\\
8.65744036384737	0.00503258656132342\\
8.09711868907005	0.00979758641066475\\
5.99563108181371	0.0228762637504772\\
4.45257316761243	0.0300452011722592\\
2.27166407033345	0.0382673640700557\\
2.16776341617934	0.0385860807078977\\
1.67326087069726	0.0387033207318202\\
};
\addlegendentry{$N =20$}

\addplot [color=mycolor3, dotted, line width=2.0pt, mark=square, mark options={solid, mycolor3}]
  table[row sep=crcr]{%
1.83746579456681	0.0276333556833912\\
0	0.0276333556833912\\
0	0\\
8.81770456115495	0\\
8.81770456115495	3.7090329843289e-09\\
8.81076099034007	0.000692022846241925\\
8.76252321046823	0.0026134586736102\\
8.72405262673134	0.00338528337241505\\
8.66910163543863	0.00412499790973992\\
8.63636424982091	0.00443320983809654\\
8.49104009736616	0.00532943743607748\\
6.632727264277	0.0136498550235196\\
3.97653834627339	0.022619911417612\\
2.52017820134887	0.0269919450432153\\
2.36966502343278	0.0273353671781562\\
2.26462330968326	0.0274987365111921\\
1.83746579456681	0.0276333556833912\\
};
\addlegendentry{$N =40$}

\addplot [color=mycolor4, dashdotted, line width=2.0pt, mark=x, mark options={solid, mycolor4}]
  table[row sep=crcr]{%
1.92697586132023	0.0156929341853527\\
0	0.0156929341853527\\
0	0\\
8.8183272162648	0\\
8.8183272162648	3.59260549629013e-09\\
8.79705312747263	0.0011834340984776\\
8.71449510548813	0.00289050386525025\\
8.6694989698604	0.00334507596844779\\
8.61820727629173	0.00370001890167476\\
8.58563007809167	0.00385293178365339\\
8.07998156196288	0.00534682829297022\\
5.73983931093962	0.0105561822763253\\
4.02882077928576	0.0135099872091105\\
3.51115343811173	0.0143127975860811\\
3.14578535673428	0.0148283101617506\\
2.91655857645647	0.0151149267525588\\
2.72534244455508	0.0153412126648229\\
2.51360524723683	0.0155222047472113\\
2.33087009972618	0.0156260932199399\\
1.92697586132023	0.0156929341853527\\
};
\addlegendentry{$N =80$}

\end{axis}
\end{tikzpicture}%
				}
			}
			\\
			\subfloat[Average Noise Power\label{fg:region_noise}]{
				\resizebox{0.48\columnwidth}{!}{
%
%
\definecolor{mycolor1}{rgb}{0.00000,0.44706,0.74118}%
\definecolor{mycolor2}{rgb}{0.85098,0.32549,0.09804}%
\definecolor{mycolor3}{rgb}{0.92941,0.69412,0.12549}%
\definecolor{mycolor4}{rgb}{0.49412,0.18431,0.55686}%
\begin{tikzpicture}

\begin{axis}[%
width=4.079in,
height=3.432in,
at={(0.684in,0.463in)},
scale only axis,
xmin=0,
xmax=12.1536195014606,
xlabel style={font=\color{white!15!black}},
xlabel={Primary Rate [bits/s/Hz]},
ymin=0,
ymax=1.61413848802014,
ylabel style={font=\color{white!15!black}},
ylabel={Total Backscatter Rate [bits/BB]},
axis background/.style={fill=white},
xmajorgrids,
ymajorgrids,
legend style={legend cell align=left, align=left, draw=white!15!black},
title style={font=\huge},
label style={font=\huge},
ticklabel style={font=\LARGE},
legend style={font=\LARGE},
scaled y ticks=false,
y tick label style={/pgf/number format/.cd, fixed, precision=2}
]
\addplot [color=mycolor1, line width=2.0pt, mark=o, mark options={solid, mycolor1}]
  table[row sep=crcr]{%
1.94323657578383	0.0912736064443333\\
0	0.0912736064443333\\
0	0\\
2.48823634841339	0\\
2.48823634841339	2.18665133304746e-08\\
2.48727504752703	0.00339348181050874\\
2.47116510401099	0.0279654977920063\\
2.45801560096861	0.0382881582438014\\
2.43637620618195	0.0493868491274363\\
2.4188690965127	0.0557228283901977\\
2.39518655866098	0.06224877080237\\
2.36636508788234	0.0682417695304301\\
2.33407448799736	0.0731094847333493\\
2.27619452201233	0.0815914771076112\\
2.2538319039682	0.0844988678149906\\
2.20591144567994	0.0880880945674867\\
1.94323657578383	0.0912736064443333\\
};
\addlegendentry{$\sigma_n^2 =-20$ dBm}

\addplot [color=mycolor2, dashed, line width=2.0pt, mark=+, mark options={solid, mycolor2}]
  table[row sep=crcr]{%
2.36575734946381	0.286606787484281\\
0	0.286606787484281\\
0	0\\
5.54607532709139	0\\
5.54607532709139	4.76230863097795e-08\\
5.54462838262602	0.00517439590429018\\
5.52015500362565	0.0422417593281575\\
5.49965524818751	0.0583166491296324\\
5.4636636308586	0.0771981220364048\\
5.434687829407	0.0885754929447449\\
5.39285970715085	0.101493000231892\\
5.32888238736008	0.116761319880763\\
5.10631884873035	0.151583224480106\\
4.82547173062891	0.181345245679938\\
4.33637644444114	0.218014267739763\\
3.9998935297534	0.238220418718349\\
3.60716249201378	0.257219285207158\\
3.24787696026814	0.271635241862735\\
2.96841605670134	0.280329500469697\\
2.36575734946381	0.286606787484281\\
};
\addlegendentry{$\sigma_n^2 =-30$ dBm}

\addplot [color=mycolor3, dotted, line width=2.0pt, mark=square, mark options={solid, mycolor3}]
  table[row sep=crcr]{%
1.6664425499052	0.80090358197302\\
0	0.80090358197302\\
0	0\\
8.83664905210358	0\\
8.83664905210358	7.58794191583025e-08\\
8.83509808785498	0.00553528024134535\\
8.80950929682915	0.0446172958945177\\
8.78741015063903	0.062034935476601\\
8.7475360265799	0.082766821355542\\
8.71578706043915	0.0943959853411858\\
8.66553087585434	0.108051088819958\\
8.10542380602146	0.20374072680512\\
5.72004167130293	0.500562562519152\\
4.18779667979098	0.644567119533987\\
2.28165934138051	0.79204182764623\\
2.17892480427208	0.798535047857861\\
1.6664425499052	0.80090358197302\\
};
\addlegendentry{$\sigma_n^2 =-40$ dBm}

\addplot [color=mycolor4, dashdotted, line width=2.0pt, mark=x, mark options={solid, mycolor4}]
  table[row sep=crcr]{%
2.21272910612094	1.61413848802014\\
0	1.61413848802014\\
0	0\\
12.1536195014606	0\\
12.1536195014606	1.04018535018589e-07\\
12.152095243709	0.00542378521360845\\
12.1267266331021	0.044210637723212\\
12.1047043868918	0.061203771216007\\
12.0649432105365	0.0809505046194507\\
12.0350239156692	0.0911375360605834\\
11.705457377251	0.165349230913179\\
7.81554157122354	0.877822002676841\\
3.78808299421503	1.44572496110324\\
2.66572914663552	1.59973188441192\\
2.56780158617495	1.61037644400617\\
2.53164869827702	1.61254250861876\\
2.21272910612094	1.61413848802014\\
};
\addlegendentry{$\sigma_n^2 =-50$ dBm}

\end{axis}
\end{tikzpicture}%
				}
			}
			\subfloat[Imperfect \gls{csi}\label{fg:region_csi}]{
				\resizebox{0.48\columnwidth}{!}{
%
%
\definecolor{mycolor1}{rgb}{0.00000,0.44706,0.74118}%
\definecolor{mycolor2}{rgb}{0.85098,0.32549,0.09804}%
\definecolor{mycolor3}{rgb}{0.92941,0.69412,0.12549}%
\definecolor{mycolor4}{rgb}{0.49412,0.18431,0.55686}%
\begin{tikzpicture}

\begin{axis}[%
width=4.079in,
height=3.432in,
at={(0.684in,0.463in)},
scale only axis,
xmin=0,
xmax=8.83442417679897,
xlabel style={font=\color{white!15!black}},
xlabel={Primary Rate [bits/s/Hz]},
ymin=0,
ymax=0.803737766460662,
ylabel style={font=\color{white!15!black}},
ylabel={Total Backscatter Rate [bits/BB]},
axis background/.style={fill=white},
xmajorgrids,
ymajorgrids,
legend style={legend cell align=left, align=left, draw=white!15!black},
title style={font=\huge},
label style={font=\huge},
ticklabel style={font=\LARGE},
legend style={font=\LARGE},
scaled y ticks=false,
y tick label style={/pgf/number format/.cd, fixed, precision=2}
]
\addplot [color=mycolor1, line width=2.0pt, mark=o, mark options={solid, mycolor1}]
  table[row sep=crcr]{%
1.6910099054664	0.803737766460662\\
0	0.803737766460662\\
0	0\\
8.83442417679897	0\\
8.83442417679897	7.5653569056105e-07\\
8.83231867904733	0.00751751725611285\\
8.80461050656468	0.0508633061046012\\
8.78255256228187	0.0685235705074172\\
8.74183742045257	0.0896342484491029\\
8.70954426290985	0.101196477198068\\
8.65241269343705	0.116180151763118\\
8.08754970249077	0.214415467044904\\
5.83255054513924	0.495500728233585\\
4.25705841240893	0.64352590860324\\
3.22038687949657	0.723685407838174\\
2.2983846383418	0.792918622258748\\
2.18800472645727	0.79933894231638\\
1.6910099054664	0.803737766460662\\
};
\addlegendentry{$\iota =0$}

\addplot [color=mycolor2, dashed, line width=2.0pt, mark=+, mark options={solid, mycolor2}]
  table[row sep=crcr]{%
1.64466773440046	0.760035568718933\\
0	0.760035568718933\\
0	0\\
8.83334506828205	0\\
8.83334506828205	1.63036842482533e-06\\
8.83119882928758	0.00742871468234786\\
8.80332510082731	0.0506857259825842\\
8.78089505204587	0.0681676002462056\\
8.73979473452631	0.0888859798933564\\
8.70750718322616	0.100370068974823\\
8.64449504663819	0.116545290185732\\
8.03361497110983	0.213267230530932\\
5.58823635238811	0.484891470161212\\
4.01397368882593	0.617471306098512\\
2.24339517524649	0.748608921549967\\
2.13411622166587	0.75519353750423\\
1.64466773440046	0.760035568718933\\
};
\addlegendentry{$\iota =0.05$}

\addplot [color=mycolor3, dotted, line width=2.0pt, mark=square, mark options={solid, mycolor3}]
  table[row sep=crcr]{%
1.70869285532576	0.715808503478469\\
0	0.715808503478469\\
0	0\\
8.83127826947351	0\\
8.83127826947351	3.33880184473467e-06\\
8.82931809241901	0.00705997191270983\\
8.8011407614724	0.0508343846182438\\
8.77855062839578	0.0684396854178604\\
8.73768445349154	0.0890392957971512\\
8.70506063710828	0.100369803053394\\
8.64585765987197	0.114635703047525\\
7.92649055684501	0.218048017278456\\
5.33825394743852	0.475052044688384\\
3.92714217649523	0.586026981343679\\
2.30014720502174	0.703125785574612\\
2.19068252594368	0.710209544891159\\
1.70869285532576	0.715808503478469\\
};
\addlegendentry{$\iota =0.1$}

\addplot [color=mycolor4, dashdotted, line width=2.0pt, mark=x, mark options={solid, mycolor4}]
  table[row sep=crcr]{%
1.66619868159017	0.638415393910433\\
0	0.638415393910433\\
0	0\\
8.82823534674063	0\\
8.82823534674063	2.39878525176457e-06\\
8.8261197574455	0.00758343262500842\\
8.79876696991114	0.0499416121710518\\
8.77610015915849	0.0673016634280347\\
8.7353518794766	0.0872005300161368\\
8.70368149759064	0.0978502670728901\\
8.62547019759099	0.112781677830968\\
7.79458194538612	0.212471933823545\\
5.1117959672575	0.434831088743396\\
2.11877703266619	0.630685292265638\\
1.66619868159017	0.638415393910433\\
};
\addlegendentry{$\iota =0.2$}

\end{axis}
\end{tikzpicture}%
				}
			}
			\\
			\subfloat[Primary \gls{snr}\label{fg:region_snr_primary}]{
				\resizebox{0.48\columnwidth}{!}{
%
%
\definecolor{mycolor1}{rgb}{0.00000,0.44706,0.74118}%
\definecolor{mycolor2}{rgb}{0.85098,0.32549,0.09804}%
\definecolor{mycolor3}{rgb}{0.92941,0.69412,0.12549}%
\definecolor{mycolor4}{rgb}{0.49412,0.18431,0.55686}%
\begin{tikzpicture}

	\begin{axis}[%
			width=4.079in,
			height=3.432in,
			at={(0.684in,0.463in)},
			scale only axis,
			xmin=0,
			xmax=9.92332189338346,
			xlabel style={font=\color{white!15!black}},
			xlabel={Primary Rate [bits/s/Hz]},
			ymin=0,
			ymax=1.71834730422335,
			ylabel style={font=\color{white!15!black}},
			ylabel={Total Backscatter Rate [bits/BB]},
			axis background/.style={fill=white},
			xmajorgrids,
			ymajorgrids,
			legend style={legend cell align=left, align=left, draw=white!15!black},
			title style={font=\huge},
			label style={font=\huge},
			ticklabel style={font=\LARGE},
			legend style={font=\LARGE},
			scaled y ticks=false,
			y tick label style={/pgf/number format/.cd, fixed, precision=2}
		]
		\addplot [color=mycolor1, line width=2.0pt, mark=o, mark options={solid, mycolor1}]
		table[row sep=crcr]{%
				1.85247429168287	1.71834730422335\\
				0	1.71834730422335\\
				0	0\\
				3.78684052274749	0\\
				3.78684052274749	1.09074571890467e-07\\
				3.72234367513494	0.241158873923192\\
				3.16679965429821	1.14892143901781\\
				2.84883713806127	1.40727728535061\\
				2.53257994128885	1.57740826228066\\
				2.38853678557276	1.63104022969941\\
				2.2565003504959	1.66887440257207\\
				2.13823977918411	1.69364619619498\\
				2.03169655873089	1.70862679617922\\
				1.98302330220642	1.71315442488467\\
				1.93644811032135	1.71622943138158\\
				1.91872656697149	1.71706060650144\\
				1.90149673316119	1.71768245560854\\
				1.88480520107176	1.71802065847498\\
				1.86828151015598	1.71826879403074\\
				1.85247429168287	1.71834730422335\\
			};
		\addlegendentry{$\gamma_\text{P} =0$ dB}

		\addplot [color=mycolor2, dashed, line width=2.0pt, mark=+, mark options={solid, mycolor2}]
		table[row sep=crcr]{%
				2.71467108243765	1.5690611320315\\
				0	1.5690611320315\\
				0	0\\
				4.90553860933898	0\\
				4.90553860933898	1.74101985141683e-07\\
				4.86633687400433	0.140300509501269\\
				4.41900651909845	0.846961386844779\\
				4.06222590319286	1.13090039213493\\
				3.61284041820999	1.37046719056819\\
				3.39923634150954	1.4511202796632\\
				3.21403602866227	1.504862867065\\
				3.05695766531175	1.53822724341412\\
				2.92384939211783	1.55729018996837\\
				2.86498591172433	1.5629109884803\\
				2.81069019335922	1.56652241037219\\
				2.79019073112918	1.5674847086388\\
				2.77029253613589	1.56820079649427\\
				2.75104389631163	1.56869140611587\\
				2.73241069049525	1.56897269695754\\
				2.71467108243765	1.5690611320315\\
			};
		\addlegendentry{$\gamma_\text{P} =10$ dB}

		\addplot [color=mycolor3, dotted, line width=2.0pt, mark=square, mark options={solid, mycolor3}]
		table[row sep=crcr]{%
				6.13643502916554	0.550186823979828\\
				0	0.550186823979828\\
				0	0\\
				7.07004431384186	0\\
				7.07004431384186	5.04984237402795e-07\\
				7.06762608009238	0.00810425421005326\\
				6.91631113074837	0.227651119452076\\
				6.75836620620173	0.352412731396597\\
				6.55958044250585	0.457999229450031\\
				6.46197216362381	0.494614192418989\\
				6.37487136177558	0.519696613767362\\
				6.30040731938322	0.535467022748589\\
				6.23688560384776	0.544551268787556\\
				6.2086301769725	0.547244394343574\\
				6.18244534615828	0.548976105487367\\
				6.17256603078898	0.549436423249413\\
				6.16298584151283	0.549779203935751\\
				6.15369231040524	0.550013227191073\\
				6.1447149885516	0.550146373234086\\
				6.13643502916554	0.550186823979828\\
			};
		\addlegendentry{$\gamma_\text{P} =20$ dB}

		\addplot [color=mycolor4, dashdotted, line width=2.0pt, mark=x, mark options={solid, mycolor4}]
		table[row sep=crcr]{%
				9.64822480763881	0.0856801173360131\\
				0	0.0856801173360131\\
				0	0\\
				9.92332189338346	0\\
				9.92332189338346	8.95272247472359e-07\\
				9.92328158136014	8.09154700357446e-05\\
				9.91587898609221	0.00944298769692819\\
				9.89607327582226	0.025036835183679\\
				9.85259421992854	0.0477177043579155\\
				9.82196151992848	0.0591435832018186\\
				9.78684530197915	0.069167352279973\\
				9.74940246593168	0.0769728814727595\\
				9.71230121082813	0.0821805038459397\\
				9.69463886223502	0.0838321813013826\\
				9.67785389651924	0.0849198151935604\\
				9.671463835881	0.0852098739217015\\
				9.66525894473783	0.0854257800416892\\
				9.65924762143605	0.0855726946386796\\
				9.65347109773093	0.0856555898417966\\
				9.64822480763881	0.0856801173360131\\
			};
		\addlegendentry{$\gamma_\text{P} =30$ dB}

	\end{axis}
\end{tikzpicture}%
				}
			}
			\subfloat[Backscatter \gls{snr}\label{fg:region_snr_backscatter}]{
				\resizebox{0.48\columnwidth}{!}{
%
%
\definecolor{mycolor1}{rgb}{0.00000,0.44706,0.74118}%
\definecolor{mycolor2}{rgb}{0.85098,0.32549,0.09804}%
\definecolor{mycolor3}{rgb}{0.92941,0.69412,0.12549}%
\definecolor{mycolor4}{rgb}{0.49412,0.18431,0.55686}%
\begin{tikzpicture}

	\begin{axis}[%
			width=4.079in,
			height=3.432in,
			at={(0.684in,0.463in)},
			scale only axis,
			xmin=0,
			xmax=8.21430854761574,
			xlabel style={font=\color{white!15!black}},
			xlabel={Primary Rate [bits/s/Hz]},
			ymin=0,
			ymax=1.85992357459342,
			ylabel style={font=\color{white!15!black}},
			ylabel={Total Backscatter Rate [bits/BB]},
			axis background/.style={fill=white},
			xmajorgrids,
			ymajorgrids,
			legend style={legend cell align=left, align=left, draw=white!15!black},
			title style={font=\huge},
			label style={font=\huge},
			ticklabel style={font=\LARGE},
			legend style={font=\LARGE},
			scaled y ticks=false,
			y tick label style={/pgf/number format/.cd, fixed, precision=2}
		]
		\addplot [color=mycolor1, line width=2.0pt, mark=o, mark options={solid, mycolor1}]
		table[row sep=crcr]{%
				6.44725641749923	0.00825221282441818\\
				0	0.00825221282441818\\
				0	0\\
				6.52524524262144	0\\
				6.52524524262144	6.36984046577934e-07\\
				6.52511577632124	0.000141554640195365\\
				6.52494378352269	0.000271993879083585\\
				6.52411403796041	0.000670458318418985\\
				6.52252597235699	0.00123674744744557\\
				6.51874573015298	0.00226949616945423\\
				6.51074503324696	0.00386802190474683\\
				6.49560678801124	0.00589689704316441\\
				6.48492459710706	0.00685925841392101\\
				6.47228501770775	0.00764219429373445\\
				6.46699275201718	0.00787045115777223\\
				6.46167023507642	0.00804561818478987\\
				6.45642676678617	0.00816617011388671\\
				6.45139658707654	0.00823350591683185\\
				6.44725641749923	0.00825221282441818\\
			};
		\addlegendentry{$\gamma_\text{B} =-20$ dB}

		\addplot [color=mycolor2, dashed, line width=2.0pt, mark=+, mark options={solid, mycolor2}]
		table[row sep=crcr]{%
				6.42375412345092	0.0758914753465313\\
				0	0.0758914753465313\\
				0	0\\
				6.68201761475096	0\\
				6.68201761475096	6.06647122640634e-07\\
				6.68193678959583	0.000247072714349278\\
				6.6768685002139	0.00671276744368922\\
				6.66059636260034	0.019450332575155\\
				6.62243320896833	0.0393648427879511\\
				6.59389885872934	0.0500419479946411\\
				6.56033839302831	0.0596429725464297\\
				6.52387302533948	0.0672612475264039\\
				6.48725977938018	0.0724108091974971\\
				6.46976330615714	0.0740521146150458\\
				6.45316911975957	0.0751320048065465\\
				6.44681920255803	0.0754213815178291\\
				6.44065406426208	0.0756369929127145\\
				6.43468782819772	0.0757837125485934\\
				6.42893673472027	0.0758667635183516\\
				6.42375412345092	0.0758914753465313\\
			};
		\addlegendentry{$\gamma_\text{B} =-10$ dB}

		\addplot [color=mycolor3, dotted, line width=2.0pt, mark=square, mark options={solid, mycolor3}]
		table[row sep=crcr]{%
				6.22682752701347	0.524033576450723\\
				0	0.524033576450723\\
				0	0\\
				7.12278253257655	0\\
				7.12278253257655	5.16303151758159e-07\\
				7.12066672673348	0.00700486506164334\\
				6.97707515187818	0.214442281390311\\
				6.82847278197181	0.332898378393172\\
				6.63539245444809	0.435259951979859\\
				6.5420356642679	0.470332192266869\\
				6.45846013363598	0.494447953364535\\
				6.38614689398235	0.509738822623144\\
				6.32431785274704	0.5185647225286\\
				6.29698128993436	0.521173255229958\\
				6.27157652383734	0.522856103371851\\
				6.26190100096475	0.523305299477758\\
				6.25259821740313	0.523638108258084\\
				6.2435836231981	0.523865160627578\\
				6.23488889053722	0.523994260605531\\
				6.22682752701347	0.524033576450723\\
			};
		\addlegendentry{$\gamma_\text{B} =0$ dB}

		\addplot [color=mycolor4, dashdotted, line width=2.0pt, mark=x, mark options={solid, mycolor4}]
		table[row sep=crcr]{%
				5.32895797753852	1.85992357459342\\
				0	1.85992357459342\\
				0	0\\
				8.21430854761574	0\\
				8.21430854761574	2.91164173554729e-07\\
				8.17344059213623	0.147335549977774\\
				7.69756724096485	0.896025760702629\\
				7.25957458328423	1.24131455006827\\
				6.63479144018421	1.57368308154365\\
				6.33103541043344	1.68855483598363\\
				6.06541470926613	1.76574273269775\\
				5.83663157486451	1.81435199972312\\
				5.64142910990468	1.84235330469723\\
				5.55432971729481	1.85071158914573\\
				5.47351667692754	1.85610606008128\\
				5.44287150461302	1.8575474718062\\
				5.41302187874583	1.85862597427057\\
				5.38408952528413	1.85936554231935\\
				5.35597043183404	1.85979019305613\\
				5.32895797753852	1.85992357459342\\
			};
		\addlegendentry{$\gamma_\text{B} =10$ dB}

	\end{axis}
\end{tikzpicture}%
				}
			}
			\caption{
				Average primary-total-backscatter rate regions for different system configurations.
			}
			\label{fg:region_config}
		\end{figure}
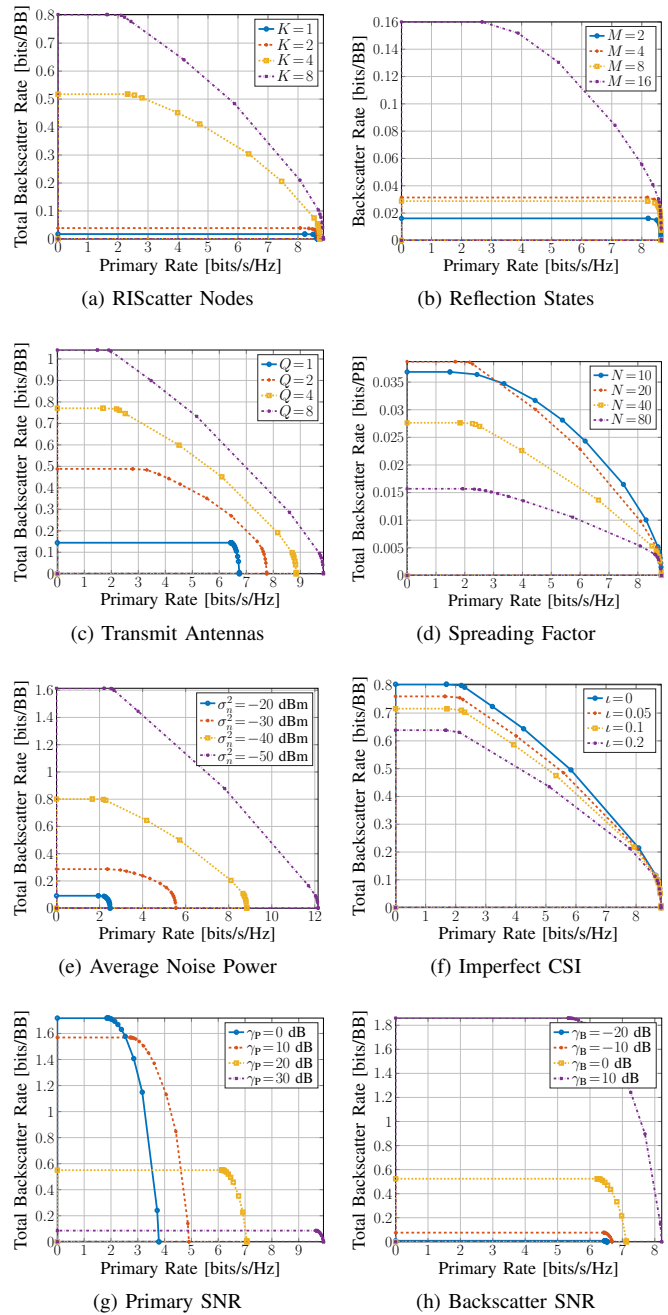
		In this study, we choose $Q=4$, $K=8$, $M=2$, $N=20$, $\sigma_v^2=\qty{-40}{dBm}$ and $r=\qty{2}{m}$ as a reference, unless otherwise specified.
		\begin{subsubsection}{Number of Nodes}
			Fig.~\subref*{fg:region_tags} reveals how the number RIScatter nodes $K$ influences the primary-backscatter tradeoff.
			Interestingly, we observe that increasing $K$ has a larger benefit on the total backscatter rate than primary.
			This is because each RIScatter node not only affects the primary \gls{snr} but also influences the relative energy difference that other nodes can make.
			To maximize the total backscatter rate, some nodes closer to the user may need to sacrifice their own rate and use the state that \emph{minimizes} the composite channel strength, in order to increase the backscatter rate of other nodes.
			This accounts for the significant primary rate decrease in the low-$\rho$ regime.
			On the other hand, when the primary link is prioritized, the RIScatter nodes boil down to \gls{ris} elements and enjoy a passive array gain of $K^2$.
		\end{subsubsection}

		\begin{subsubsection}{Number of States}
			Fig.~\subref*{fg:region_states} shows the relationship between the available reflection states (i.e., \gls{qam} order) $M$ and the achievable rate region when $K=1$.
			We notice that increasing the reflection states has a marginal effect on the primary rate but significantly improves the backscatter rate.
			This is because the maximum amplitude normalized-\gls{qam} \eqref{eq:backscatter_modulation} involves more weak reflection points as $M$ increases.
			It enhances the receive energy diversity but cannot provide enough phase shift resolution with maximum reflection.
		\end{subsubsection}

		\begin{subsubsection}{Number of Transmit Antennas}
			Fig.~\subref*{fg:region_txs} illustrates the impact of transmit antennas $Q$ on the average performance.
			As $Q$ increases, more scattered paths become available and the channel diversity can be better exploited to improve the primary-backscatter tradeoff.
			It emphasizes the importance of multi-antenna RIScatter systems and demonstrate the effectiveness of the proposed \gls{pga} design.
		\end{subsubsection}

		\begin{subsubsection}{Spreading factor}
			Fig.~\subref*{fg:region_duration} shows how the spreading factor $N$ affects the achievable rate region.\footnote{Here, the unit of total backscatter rate is bits per primary block to show backscatter throughput.}
			Using a very large $N$ (as in the case of \gls{sr}) can severely constrain the backscatter throughput, since the gain in energy certainty (by the law of large numbers) cannot withstand the loss in the gross rate.
			As $N \to \infty$, RIScatter nodes boil down to static \gls{ris} elements and the total backscatter rate approaches \num{0}.
			On the other hand, when $N$ is too small, the \gls{dmmac} \eqref{eq:dmmac} becomes unreliable and energy detection is error-prone.
			It explains the observation that $N=10$ provides lower backscatter throughput than $N=20$.
			Therefore, we conclude the spreading factor $N$ should be carefully designed over multiple factors (e.g., primary and backscatter \gls{snr}, data rate requirements, load switching speed at the nodes, and signal processing capability at the user).
		\end{subsubsection}

		\begin{subsubsection}{Average Noise Power}
			Fig.~\subref*{fg:region_noise} depicts the impact of average noise power $\sigma_v^2$ on average rate regions.
			Note that the noise influences both primary and backscatter \gls{snr}.
			When $\sigma_v^2$ relatively high, one can choose a larger $N$ to improve the \gls{snr} of energy detection.
		\end{subsubsection}


		\begin{subsubsection}{Imperfect \gls{csi}}
			Due to the lack of \gls{rf} chains at RIScatter nodes, fast and accurate acquisition of the cascaded \gls{csi} can be challenging, especially when the backscatter \gls{snr} is weak or the number of nodes is large.
			We consider an imperfect \gls{csi} model, where the cascaded channel of node $k$ is estimated as
			\begin{equation}
				\hat{\boldsymbol{h}}_{\text{C},k} = \boldsymbol{h}_{\text{C},k} + \tilde{\boldsymbol{h}}_{\text{C},k},
			\end{equation}
			$\tilde{\boldsymbol{h}}_{\text{C},k}$ is the estimation error with entries following i.i.d. CSCG distribution $\mathcal{CN}(0, \iota \Lambda_\text{C})$, $\iota$ is the relative estimation error, and $\Lambda_\text{C}$ is the cascaded path loss.
			In Fig.~\subref*{fg:region_csi}, it is observed that the channel estimation error mainly affects the backscatter rate.
			When $\iota$ increases from \num{0} to \num{0.2}, the maximum total backscatter rate decreases by \qty{20}{\percent}.
			This is because the energy detector is sensitive to the \gls{dmmac} \eqref{eq:dmmac} and thus the estimation of the cascaded channel.
			On the other hand, a small estimation error may be insufficient to change the optimal passive beamforming state and the primary rate is almost unchanged.
		\end{subsubsection}

		\begin{subsubsection}{Primary \gls{snr}}
			\label{sc:primary_snr}
			The backscatter \gls{snr} is fixed to \qty{0}{dB} with $Q=1$ in this study.
			Interestingly, Fig.~\subref*{fg:region_snr_primary} shows that increasing the primary \gls{snr} can improve the primary rate but degrade the backscatter rate.
			The reason is that the relative strength of the scattered signal compared to the direct signal is weakened, such that the nodes cannot make enough difference to the energy detector.
			This, together with Fig.~\subref*{fg:region_beamforming} and \subref*{fg:region_tags}, emphasizes the importance of balancing the primary and backscatter \gls{snr} in the design of active-passive coexisting networks.
		\end{subsubsection}

		\begin{subsubsection}{Backscatter \gls{snr}}
			The primary \gls{snr} is fixed to \qty{20}{dB} with $Q=1$ in this study.
			Fig.~\subref*{fg:region_snr_backscatter} shows that the primary and backscatter rates are both improved when the backscatter \gls{snr} increases.
			This motivates one to use high-efficiency or semi-passive RIScatter nodes to improve the overall performance.
			In a multi-user RIScatter network, each node may be assigned to the nearest user to guarantee uniformly good performance for both links.
		\end{subsubsection}
	\end{subsection}
\end{section}

\begin{section}{Conclusion}
	\label{sc:conclusion}
	This paper introduced RIScatter as a novel scatter protocol that bridges backscatter modulation and passive beamforming.
	Starting from scattering principles, we showed how RIScatter nodes generalize information nodes of \gls{bc} and reflect elements of \gls{ris}, how they can be built over existing passive scatter devices, and how they simultaneously encode self information and assist legacy transmission.
	We also proposed a practical \gls{sic}-free receiver that exploits the properties of active-passive coexisting networks to benefit both subsystems.
	The achievable primary-total-backscatter rate region was then studied for a single-user multi-node RIScatter network, where the input distribution, active beamforming, and decision thresholds are iteratively updated.
	Numerical results validated the proposed algorithms and emphasized the importance of adaptive input distribution and cooperative receiver design.

	To support massive RIScatter networks with large number of nodes and states, one possible future direction is to consider backscatter detection over the received signal domain rather than energy domain, where multi-antenna \cite{Liu2022c} and learning-based approaches can be promising.
	Another interesting question is how to design RIScatter nodes and receivers in a multi-user system to fully exploit the dynamic passive beamforming that naturally origins from backscatter modulation.
\end{section}

\begin{appendix}
	\begin{subsection}{Proof of Proposition \ref{pr:input_kkt_condition}}
		Denote the Lagrange multipliers associated with \eqref{co:sum_probability} and \eqref{co:nonnegative_probability} as $\{\nu_k\}_{k \in \mathcal{K}}$ and $\{\lambda_{m_k}\}_{k \in \mathcal{K},m_k \in \mathcal{M}}$, respectively.
		The Lagrangian function of problem \eqref{op:input_distribution} is
		\begin{equation}
			L(p, \nu, \lambda) = I(x_{\mathcal{K}}) + \sum_k \nu_k \Bigl( \sum_{m_k} p(x_{m_k}) - 1 \Bigr) + \sum_{k,m_k} \lambda_{m_k} p(x_{m_k}),
		\end{equation}
		and the \gls{kkt} conditions are, $\forall k,m_k$,
		\begin{subequations}
			\label{eq:input_kkt_condition_original}
			\begin{gather}
				I_k^\star(x_{m_k}) - (1 - \rho) + \nu_k^\star + \lambda_{m_k}^\star = 0, \label{eq:input_kkt_condition_1} \\
				\lambda_{m_k}^\star = 0, \quad    \text{if} \ p^\star(x_{m_k}) > 0, \label{eq:input_kkt_condition_2} \\
				\lambda_{m_k}^\star \ge 0, \quad  \text{if} \ p^\star(x_{m_k}) = 0 \label{eq:input_kkt_condition_3}.
			\end{gather}
		\end{subequations}
		Plugging \eqref{eq:input_kkt_condition_2} and \eqref{eq:input_kkt_condition_3} into \eqref{eq:input_kkt_condition_1} yields
		\begin{subequations}
			\label{eq:input_kkt_condition_transformed}
			\begin{alignat}{2}
				I_k^\star(x_{m_k}) & = 1 - \rho - \nu_k^\star, \quad   &  & \text{if} \ p^\star(x_{m_k}) > 0,\label{eq:probable_states_marginal} \\
				I_k^\star(x_{m_k}) & \le 1 - \rho - \nu_k^\star, \quad &  & \text{if} \ p^\star(x_{m_k}) = 0,\label{eq:dropped_states_marginal}
			\end{alignat}
		\end{subequations}
		such that
		\begin{equation}
			\sum_{m_k} p^\star(x_{m_k}) I_k^\star(x_{m_k}) = 1 - \rho - \nu_k^\star.
			\label{eq:input_kkt_condition_implied}
		\end{equation}
		On the other hand, by definition \eqref{eq:weighted_sum_marginal_information} we have
		\begin{equation}
			\sum_{m_k} p^\star(x_{m_k}) I_k^\star(x_{m_k}) = I^\star(x_{\mathcal{K}}),
			\label{eq:weighted_sum_marginal_information_implied}
		\end{equation}
		where the right-hand side is irrelevant to $k$.
		\eqref{eq:input_kkt_condition_transformed}, \eqref{eq:input_kkt_condition_implied}, and \eqref{eq:weighted_sum_marginal_information_implied} together complete the proof.
		\label{ap:input_kkt_condition}
	\end{subsection}

	\begin{subsection}{Proof of Proposition \ref{pr:input_kkt_solution}}
		We first prove sequence \eqref{eq:input_kkt_solution} is non-decreasing in weighted sum mutual information.
		Let $p(x_{m_{\mathcal{K}}}) = \prod_{q \in \mathcal{K}} p(x_{m_q})$ and $p'(x_{m_{\mathcal{K}}}) = p'(x_{m_k}) \prod_{q \in \mathcal{K} \setminus \{k\}} p(x_{m_q})$ be two distributions potentially different at $x_{m_k}$,
		and $J \bigl(p(x_{m_{\mathcal{K}}}),p'(x_{m_{\mathcal{K}}}) \bigr)$ be a joint function defined in \eqref{eq:intermediate_information_function} at the end of page \pageref{eq:intermediate_information_function}.
		\begin{figure*}[!b]
			\hrule
			\begin{equation}
				J \bigl( p(x_{m_{\mathcal{K}}}),p'(x_{m_{\mathcal{K}}}) \bigr) \triangleq \sum_{m_{\mathcal{K}}} p(x_{m_{\mathcal{K}}})
				\Biggl( \rho \log \Bigl(1 + \frac{\lvert \boldsymbol{h}^\mathsf{H}(x_{m_{\mathcal{K}}}) \boldsymbol{w} \rvert^2}{\sigma_v^2}\Bigr) + (1 - \rho) \sum_{m_{\mathcal{K}}'} q(\hat{x}_{m_{\mathcal{K}}'}|x_{m_{\mathcal{K}}}) \log \frac{q(\hat{x}_{m_{\mathcal{K}}'}|x_{m_{\mathcal{K}}}) p'(x_{m_{\mathcal{K}}})}{p'(\hat{x}_{m_{\mathcal{K}}'}) p(x_{m_{\mathcal{K}}})} \Biggr).
				\label{eq:intermediate_information_function}
			\end{equation}
		\end{figure*}
		It is straightforward to verify $J \bigl( p(x_{m_{\mathcal{K}}}),p(x_{m_{\mathcal{K}}}) \bigr) = I(x_{\mathcal{K}})$ and $J \bigl( p(x_{m_{\mathcal{K}}}),p'(x_{m_{\mathcal{K}}}) \bigr)$ is a concave function for a given $p'(x_{m_{\mathcal{K}}})$.
		Setting $\nabla_{p(x_{m_k})} J \bigl( p(x_{m_{\mathcal{K}}}),p'(x_{m_{\mathcal{K}}}) \bigr) = 0$ yields
		\begin{equation}
			S_k'(x_{m_k}) - S_k'(x_{i_k}) + (1 - \rho) \log \frac{p(x_{i_k})}{p^\star(x_{m_k})} = 0,
			\label{eq:optimal_intermediate_information_condition}
		\end{equation}
		where $i_k \ne m_k$ is the reference state and
		\begin{align}
			S_k'(x_{m_k})
			 & \triangleq I_k'(x_{m_k}) + (1 - \rho) \sum_{m_{\mathcal{K} \setminus \{k\}}} p(x_{m_{\mathcal{K} \setminus \{k\}}})\nonumber \\
			 & \quad \times \sum_{m_{\mathcal{K}}'} q(\hat{x}_{m_{\mathcal{K}}'}|x_{m_{\mathcal{K}}}) \log p'(x_{m_{\mathcal{K}}}).
		\end{align}
		Evidently, $\forall m_k \ne i_k$, \eqref{eq:optimal_intermediate_information_condition} boils down to
		\begin{equation}
			p^\star(x_{m_k}) = \frac{p'(x_{m_k}) \exp \Bigl( \frac{\rho}{1 - \rho} I_k'(x_{m_k}) \Bigr)}{\sum_{m_k'} p'(x_{m_k'}) \exp \Bigl( \frac{\rho}{1 - \rho} I_k'(x_{m_k'}) \Bigr)}.
			\label{eq:optimal_relative_distribution}
		\end{equation}
		Since $p(x_{i_k}) = 1 - \sum_{m_k \ne i_k} p^\star(x_{m_k})$ has exactly the same form as \eqref{eq:optimal_relative_distribution}, the choice of reference does not matter and \eqref{eq:optimal_relative_distribution} is optimal $\forall m_k \in \mathcal{M}$.
		That is, for a fixed $p'(x_{m_{\mathcal{K}}})$, \eqref{eq:optimal_relative_distribution} ensures
		\begin{equation}
			J \bigl( p(x_{m_{\mathcal{K}}}),p'(x_{m_{\mathcal{K}}}) \bigr) \ge I'(x_{\mathcal{K}}).
			\label{eq:information_difference_lower}
		\end{equation}
		On the other hand, we notice
		\begin{subequations}
			\allowdisplaybreaks
			\label{eq:information_difference_upper}
			\begin{align}
				 & I(x_{\mathcal{K}}) - J \bigl( p(x_{m_{\mathcal{K}}}),p'(x_{m_{\mathcal{K}}}) \bigr)                                                                                                                    \nonumber           \\
				 & = (1 - \rho) \sum_{m_k} \frac{p'(x_{m_k}) f_k'(x_{m_k})}{\sum_{m_k'} p'(x_{m_k'}) f_k'(x_{m_k'})} \sum_{m_{\mathcal{K}}''} q(\hat{x}_{m_{\mathcal{K}}''}|x_{m_k})                                      \nonumber           \\
				 & \quad \times \log \frac{\sum_{m_k'} p'(x_{m_k'}) q(\hat{x}_{m_{\mathcal{K}}''}|x_{m_k'}) f_k'(x_{m_k})}{\sum_{m_k'} p'(x_{m_k'}) q(\hat{x}_{m_{\mathcal{K}}''}|x_{m_k'}) f_k'(x_{m_k'})}                                   \\
				 & \ge (1 - \rho) \sum_{m_k} \frac{p'(x_{m_k}) f_k'(x_{m_k})}{\sum_{m_k'} p'(x_{m_k'}) f_k'(x_{m_k'})} \sum_{m_{\mathcal{K}}''} q(\hat{x}_{m_{\mathcal{K}}''}|x_{m_k})                                    \nonumber           \\
				 & \quad \times \Biggl( 1 - \frac{\sum_{m_k'} p'(x_{m_k'}) q(\hat{x}_{m_{\mathcal{K}}''}|x_{m_k'}) f_k'(x_{m_k'})}{\sum_{m_k'} p'(x_{m_k'}) q(\hat{x}_{m_{\mathcal{K}}''}|x_{m_k'}) f_k'(x_{m_k})} \Biggr)                    \\
				 & = (1 - \rho) \Biggl( 1 - \sum_{m_k} \frac{p'(x_{m_k}) \cancel{f_k'(x_{m_k})}}{\sum_{m_k'} p'(x_{m_k'}) f_k'(x_{m_k'})} \sum_{m_{\mathcal{K}}''} q(\hat{x}_{m_{\mathcal{K}}''}|x_{m_k})                 \nonumber           \\
				 & \quad \times \frac{\sum_{m_k'} p'(x_{m_k'}) q(\hat{x}_{m_{\mathcal{K}}''}|x_{m_k'}) f_k'(x_{m_k'})}{\sum_{m_k'} p'(x_{m_k'}) q(\hat{x}_{m_{\mathcal{K}}''}|x_{m_k'}) \cancel{f_k'(x_{m_k})}} \Biggr)                       \\
				 & = (1 - \rho) \Biggl( 1 - \sum_{m_{\mathcal{K}}''} \frac{\cancel{\sum_{m_k} p'(x_{m_k}) q(\hat{x}_{m_{\mathcal{K}}''}|x_{m_k})}}{\sum_{m_k'} p'(x_{m_k'}) f_k'(x_{m_k'})}                                        \nonumber  \\
				 & \quad \times \frac{\sum_{m_k'} p'(x_{m_k'}) q(\hat{x}_{m_{\mathcal{K}}''}|x_{m_k'}) f_k'(x_{m_k'})}{\cancel{\sum_{m_k'} p'(x_{m_k'}) q(\hat{x}_{m_{\mathcal{K}}''}|x_{m_k'})}} \Biggr)                                     \\
				 & = (1 - \rho) \Biggl( 1 - \frac{\sum_{m_k'} p'(x_{m_k'}) f_k'(x_{m_k'}) \sum_{m_{\mathcal{K}}''} q(\hat{x}_{m_{\mathcal{K}}''}|x_{m_k'})}{\sum_{m_k'} p'(x_{m_k'}) f_k'(x_{m_k'})} \Biggr)                        \nonumber \\
				 & = 0,
			\end{align}
		\end{subequations}
		where $f_k'(x_{m_k}) \triangleq \exp \bigl( \frac{\rho}{1 - \rho} I_k'(x_{m_k}) \bigr)$ and the equality holds if and only if $p(x_{m_{\mathcal{K}}})$ and $p'(x_{m_{\mathcal{K}}})$ equals (i.e., \eqref{eq:optimal_relative_distribution} converges).
		\eqref{eq:information_difference_lower} and \eqref{eq:information_difference_upper} together imply $I(x_{\mathcal{K}}) \ge I'(x_{\mathcal{K}})$.
		Since mutual information is bounded above, we conclude the sequence \eqref{eq:input_kkt_solution} is non-decreasing and convergent.

		Next, we prove any converging point of sequence \eqref{eq:input_kkt_solution}, denoted as $p^\star(x_{m_k})$, fulfills the \gls{kkt} conditions \eqref{eq:input_kkt_condition}.
		Let
		\begin{equation}
			D^{(r)}(x_{m_k}) \triangleq \frac{p^{(r+1)}(x_{m_k})}{p^{(r)}(x_{m_k})} = \frac{f_k^{(r)}(x_{m_k})}{\sum_{m_k'} p^{(r)}(x_{m_k'}) f_k^{(r)}(x_{m_k'})}.
		\end{equation}
		As sequence \eqref{eq:input_kkt_solution} is convergent, any state with $p^\star(x_{m_k}) > 0$ need to satisfy $D^\star(x_{m_k}) \triangleq \lim_{r \to \infty} D^{(r)}(x_{m_k}) = 1$, namely
		\begin{equation}
			I_k^\star(x_{m_k}) = \frac{1 - \rho}{\rho} \log \sum_{m_k'} p^\star(x_{m_k'}) \exp \Bigl( \frac{\rho}{1 - \rho} I_k^\star(x_{m_k'}) \Bigr).
		\end{equation}
		The right-hand side is a constant for node $k$ and implies \eqref{eq:probable_states_marginal}.
		That is, any converging point with nonzero probability must satisfy \eqref{eq:probable_states}.
		On the other hand, we assume $p^\star(x_{m_k}) = 0$ does not satisfy \eqref{eq:dropped_states}, namely
		\begin{equation}
			I_k^\star(x_{m_k}) > I^\star(x_{\mathcal{K}}) = \sum_{m_k'} p^\star(x_{m_k'}) I_k^\star(x_{m_k'}),
			\label{eq:dropped_states_assumption}
		\end{equation}
		Since the exponential function is monotonically increasing, \eqref{eq:dropped_states_assumption} implies $f_k^\star(x_{m_k}) > \sum_{m_k'} p^\star(x_{m_k'}) f_k^\star(x_{m_k'})$ and $D^\star(x_{m_k}) > 1$.
		It contradicts with
		\begin{equation}
			p^{(r)}(x_{m_k}) = p^{(0)}(x_{m_k}) \prod_{n=1}^r D^{(n)}(x_{m_k}),
		\end{equation}
		since the left-hand side is zero while all terms on the right-hand side are strictly positive.
		The proof is completed.
		\label{ap:input_kkt_solution}
	\end{subsection}
\end{appendix}

\begin{section}{Acknowledgement}
	The authors would like to thank Prof. Geoffrey Ye Li for his clinical comments on earlier versions of the manuscript.
\end{section}

\bibliographystyle{IEEEtran}
\bibliography{library.bib}
\end{document}